\documentclass[aps,prb,amsfonts,amssymb,twocolumn,amsmath,preprintnumbers,nofootinbib,superscriptaddress,floatfix,showpacs]{revtex4}
\usepackage{amsmath,bm,amsfonts,amssymb}
\usepackage[dvips]{graphics}
\usepackage{graphicx}

\begin{document}

\title{Influence of the atomic-scale inhomogeneity of the pair interaction
on extracted from the STM spectra characteristics of high-$T_c$
superconductors.}
\author{A. M. Bobkov}
\affiliation{Institute of Solid State Physics, Chernogolovka,
Moscow reg., 142432 Russia}
\author{I. V. Bobkova}
\email[Electronic address: ]{bobkova@issp.ac.ru}
\affiliation{Institute of Solid State Physics, Chernogolovka,
Moscow reg., 142432 Russia}

\date{\today}

\begin{abstract}
The influence of the atomic-scale inhomogeneities of the pairing
interaction strength on the superconducting order parameter and
the conductance spectra measurable by STM is studied in the
framework of weak-coupling BCS-like theory for two-dimensional
lattice model. First of all, it is found that the inhomogeneity
having the form of atomic-scale regions of enhanced pair
interaction increases the ratio of the local low-temperature gap
in differential conductance spectra to the local temperature of
vanishing the gap $2\Delta_g/T_p$. Even in the framework of
mean-field treatment this ratio is shown to be larger than the one
corresponding to the homogeneous case. It is shown that the effect
of thermal phase fluctuations of the superconducting order
parameter can further increase this ratio. Taking them into
account in the framework of a toy model we obtained the ratio
$2\Delta_g/T_p$ to be $\sim 7-8$. It is found that the additional
atomic-scale hopping element disorder and weak potential
scatterers, which can also take place in cuprate materials, have
no considerable effect on the statistical properties of the
system, including the distribution of the gaps, $T_p$ and the
ratio $2\Delta_g/T_p$. The second consequence of the atomic-scale
order parameter inhomogeneity is the anticorrelation between the
low-temperature gap and the high-temperature zero-bias
conductance. The obtained results could bear a relation to recent
STM measurements.
\end{abstract}

\maketitle

\section{Introduction}
Nanoscale inhomogeneities have been widely observed in the
high-temperature superconductor $\rm Bi_2Sr_2CaCu_2O_{8+x}$
(BSCCO) and have generated intense interest
\cite{cren00,pan01,howard01,lang02,mcelroy05,fang06}. In
particular, the spectral gap in the local density of states (LDOS)
has been investigated by scanning tunneling microscopy (STM). It
was found that deeply in the superconducting state the
low-temperature gap varies by a factor of 2 over distances of
$20-30 $ $\overset{\circ}{\rm A}$. Recently the existence of very
similar picture of inhomogeneities in $La_{2-x}Sr_xCuO_4$ has been
also reported \cite{kato08}. Several scenarios have been proposed
to understanding this electronic inhomogeneity. First of all, it
was speculated that poorly screened electrostatic potentials of
the dopant atoms vary a doping concentration locally, giving rise
to the gap modulations \cite{martin01,zwang02,qwang02,zhou06}.
Alternatively, these inhomogeneities are associated with a
competing order parameter, such as spin
\cite{kivelson03,atkinson05,alvarez05} and orbital
antiferromagnetism \cite{ghosal04}, charge density wave
\cite{podolsky03} or pair density wave \cite{chen04}. Further, the
positive correlations between the inhomogeneities and positions of
the dopant atoms have been observed by STM on the optimally doped
BSCCO \cite{mcelroy05}. After that it was proposed by Nunner {\it
et al.} in Ref.~\onlinecite{nunner05} that the dopant atoms
modulate the pairing interaction locally on the atomic scale. The
LDOS calculated in the framework of this model is in good
agreement with the key characteristics of the experimental
spectra. However, there is an alternative picture of the
inhomogeneity origin based on the local variation of the doping
concentration, which explains why the correlations are rather weak
\cite{zhou06}.

On the other hand, in the high-$T_c$ superconductors a partial gap
in the LDOS exists for a range of temperatures above $T_c$
\cite{timusk99}. There is no consensus up to now if this gap is
due to pairing without phase coherence, a competing order or
proximity to the Mott state \cite{norman05,lee06,millis06,cho06}.
The inhomogeneities described above complicate the situation. Only
very recently the spatially resolved STM measurements of gap
formation in BSCCO samples with different $T_c$ corresponding to
hole concentrations from $0.12$ to $0.22$ were performed
\cite{gomes07}. For a range of doping from $0.16$ to $0.22$ they
have found that gaps nucleate in nanoscale regions above $T_c$ and
proliferate as the temperature is lowered, evolving to the spatial
distribution of gap values in the superconducting state. It was
observed experimentally that overdoped and optimally doped samples
have identical gap-temperature scaling ratios. Taking into account
this finding together with the fact that in the overdoped samples
pseudogap effects are believed to be weak or absent and
consistency of the low-temperature spectra with a $d$-wave
superconducting gap, Gomes $\it et. al$\cite{gomes07} have
interpreted the gaps above $T_c$ as those associated with pairing.
Despite the inhomogeneity, every pairing gap develops locally at
the temperature $T_p$ following the relation $2\Delta_g/T_p=7.9
\pm 0.5$ in wide range of doping from overdoped to optimally doped
samples. This local pairing criterion seems to fail only in
underdoped samples and DOS indicates the presence of another
phenomenon, possibly unrelated to pairing.

It is well-known that in the framework of weak-coupling BCS theory
with homogeneous pairing amplitude the ratio $2\Delta_g/T_c$ is
$3.5$ for $s$-wave superconductors  and approximately is in the
interval $4.3-4.5$ for d-wave superconductors (this value depends
slightly on the particular tight-binding parameters). The ratio
$2\Delta_g/T_p \sim 4.7-5.2$ for d-wave case is a bit higher due
to thermal smearing of the measured dI/dV curves. It is worth to
note that in the framework of strong-coupling theory the ratio
$2\Delta_g/T_c$ is in the range $3.5-5$ for s-wave
superconductors, but becomes dependent on $\Delta_g$ and can reach
the values $\sim 10$ for d-wave pairing case \cite{scalapino94}.

In the present paper we show that if the superconducting order
parameter (OP) is modulated locally on the atomic scale, the ratio
$2\Delta_g/T_p$ strongly increases in the framework of
conventional weak-coupling theory. In addition, this inhomogeneity
inevitably leads to the anticorrelation between the
low-temperature gap and the high-temperature zero-bias
conductance. This findings resemble the results observed in recent
STM experiments\cite{gomes07,yazdani08}. The most natural way to
obtain the atomic-scale modulations of the OP is to assume the
atomic-scale modulations of the pairing interaction strength, as
it was proposed by Nunner {\it et. al} \cite{nunner05}. We do not
mean any particular mechanism of pairing. Our analysis is
phenomenological and the conclusions are independent on the
underlying pairing mechanism and the particular course of the
local pair interaction modulations. Here we only focus on the
effect of the inhomogeneity on some observable properties of
cuprate superconductors. As it is discussed below, if this
inhomogeneity has the form of atomic-scale regions of enhanced
pair interaction, the ratio $2\Delta_g/T_p$ increases even in the
framework of mean-field treatment. However, the effect only takes
place if the total area of these regions is less than the area
occupied by the background interaction. The last condition is
essential. If this is not the case, that is the total area of the
enhanced pairing regions is of the order of or larger than the
area of the background pairing, the discussed ratio is reduced or,
at least, remains equal to the homogeneous one. It was
demonstrated in a number of papers. In particular, the attractive
Hubbard model with inhomogeneous pairing amplitude was
studied\cite{aryanpour} and it was shown that the superconducting
critical temperature can be significantly increased in such a
system as compared to a uniform system corresponding to the
pairing interaction averaged over the system. The zero temperature
superconducting order parameter increases also, but in a less
degree than the critical temperature, what results in reducing the
ratio $2\Delta_g/T_c$. The reduction of the ratio of the energy
gap to the critical temperature due to the inhomogeneity of
coupling constant was also obtained for dirty s-wave
superconductors\cite{zou07}. The potential disorder, as was
demonstrated \cite{franz97}, also diminishes this quantity because
the superconducting order parameter is suppressed on the distance
of the order of the coherence length around an impurity and,
consequently, the total area of the enhanced order parameter
regions dominates.

Further, it is physically reasonable that the phase of the
superconducting order parameter should fluctuate from one region
of enhanced pairing amplitude to another in such an inhomogeneous
situation, especially for short coherence length cuprate
superconductors. In particular, the state with antiferromagnetic
and nano-scale superconducting domains exhibiting randomly
distributed phases was recently proposed \cite{alvarez08} to
account for the formation of the Fermi arcs, observed in the
pseudogap phase of the underdoped
cuprates\cite{norman98,kanigel06,kanigel07}. The existence of the
thermal phase fluctuations of the superconducting clustered state
in disordered s-wave superconductors and their role in the
superconductor-insulator transition was demonstrated\cite{dubi07}.
In the present paper we show that thermal phase fluctuations can
significantly suppress the temperature $T_p$ and, consequently,
increase the ratio $2\Delta_g/T_p$. Even in the framework of very
simple model we study here this quantity reaches the value $\sim
7-8$, comparable to the experimentally observed\cite{gomes07}.

The influence of the additional disorder such as the atomic-scale
inhomogeneities of the hopping matrix elements and weak potential
scatterers is also considered. It is shown that while they can
affect the shape of low-temperature LDOS in the system, the
properties of interest: gap's distribution, $T_p$'s distribution
and the distribution of the ratio $2\Delta_g/T_p$ remain
qualitatively unchanged.

The paper is organized as follows. Sec.~\ref{mf} is devoted to the
detailed mean-field consideration of the problem. The model
mean-field Hamiltonian we use and the outline of the T-matrix
method are introduced in Sec.~\ref{method}. The single pair
interaction perturbation is studied and the physical reasons for
the enhancement of the ratio $2\Delta_g/T_p$ are discussed in
Sec.~\ref{single_phys}. Sec.~\ref{many_phys} is devoted to the
mean-field treatment of the interaction between many OP
scatterers. The effect of additional weak potential and hopping
element inhomogeneities is investigated in Sec.~\ref{add_dis}. The
influence of the thermal phase fluctuations on the quantities
under consideration is discussed in Sec.~\ref{phase}. In
Sec.~\ref{anticorr_phys} it is demonstrated that the atomic-scale
OP inhomogeneity leads to the anticorrelation between the local
low-temperature gap and the value of zero-bias conductance at
higher temperatures. The conclusions are presented in
Sec.~\ref{conclusions}.

\section{Mean-field treatment}
\label{mf}
\subsection{Model and method}
\label{method}

We consider the following Hamiltonian on a square lattice
\begin{equation}
\hat H = -\sum \limits_{ij, \sigma} t_{ij} c_{i\sigma}^\dagger
c_{j \sigma}-\sum \limits_{i,\sigma}\mu c_{i\sigma}^\dagger c_{i
\sigma}+\sum \limits_{\langle ij \rangle}\left(
\Delta_{ij}c_{i\uparrow}^\dagger c_{j\downarrow}^\dagger + h.c.
\right) \label{hamiltonian} \enspace ,
\end{equation}
where $c_{i\sigma}(c_{i\sigma}^\dagger)$ stands for an electron
annihilation (creation) operator at site $i$ with spin $\sigma$.
$\sum_{ij}$ indicates summation over neighboring sites, while
$\sum_{\langle ij \rangle}$ denotes the summation over nearest
neighbors. $t_{ij}$ is the hopping integral between sites $i$ and
$j$. We set $t_{ij}$ to be $t=1$ for the nearest-neighbor hopping
and all the energies are measured in units of $t$ throughout the
paper. The nearest-neighbor $d$-wave order parameter should be
determined self-consistently: $\Delta_{ij}=-g_{ij}\langle
c_{i\downarrow}c_{j\uparrow}-c_{j\downarrow}c_{i\uparrow}
\rangle$.

In order to analyze the inhomogeneous pairing correlations in the
framework of the mean-field Hamiltonian (\ref{hamiltonian}) we
exploit the fully self-consistent T-matrix technique for Gor'kov
Green's functions. Starting from the Gor'kov equations the normal
and anomalous Green's functions are expressed in terms of the
homogeneous background Green's functions $\check G_{ij}^0$ and the
T-matrix, which contains all the inhomogeneities. Then the full
Green's function, which depends on two space indices $i$ and $j$,
takes the form
\begin{equation}
\check G_{ij}=\check G_{ij}^0 + \sum \limits_{k,m} \check G_{ik}^0
\check T_{km} \check G_{mj}^0 \label{Green_function} \enspace .
\end{equation}
Here $\check T_{km}=-\sum \limits_n (\check M^{-1})_{kn}\check
V_{nm}$, $\check M_{km}=\delta_{km} + \sum \limits_n \check
G_{kn}^0 \check V_{nm}$. $\check V_{km}$ is the perturbation
matrix including all the inhomogeneities. All Green's functions
and T-matrices are $4 \times 4$ matrices in the direct product of
spin and particle-hole spaces, what indicated by the symbol
$\check ~$. $\hat \tau_i$ and $\hat \sigma_i$ are Pauli matrices
in particle-hole and spin spaces, respectively. The summation is
taken over all the sites, where the OP
$\Delta_{km}=\Delta_{km}^0+\delta \Delta_{km}$ differs from the
background value $\Delta_{km}^0$. $\Delta_{km}^0$ is assumed to be
of $d$-wave type, that is $\Delta_{ii \pm \hat a}^0=-\Delta_{ii
\pm \hat b}^0=\Delta^0$. $\hat a$ and $\hat b $ are basis vectors
of the square lattice. We set the lattice constant $a$ to be equal
to unity.

OP is to be calculated from the self-consistency equation
\begin{equation}
\Delta_{ij} = g_{ij}T\sum \limits_{\varepsilon_n}{\rm Tr}_4 \left[
\hat \tau_- i \hat \sigma_y \check G_{ij}(\varepsilon_n) \right]
\label{self_consistency} \enspace ,
\end{equation}
where $\hat \tau_-=(\hat \tau_x - i \hat \tau_y)/2$ and
$\varepsilon_n$ is the fermionic Matsubara frequency. The energy
cutoff $|\varepsilon_n|<4$ is used in Eq.~(\ref{self_consistency})
for the technical convenience. We have checked that the particular
value of the cutoff energy (if it is considerably larger than the
maximal value of the OP) does not change qualitatively the results
and only slightly varies the OP quantitative values.
Eqs.~(\ref{self_consistency}) and (\ref{Green_function}) allow us
to find the OP $\Delta_{ij}$ numerically. Making use of the
outlined technique we consider a square of the size $n \times n$,
immersed into an infinite lattice carrying the homogeneous OP
$\Delta^0_{ij}$, as an inhomogeneity described by T-matrix.

\subsection{Single perturbation}
\label{single_phys}

In general, the T-matrix can describe inhomogeneities of all the
parameters entering the equations: hopping elements, chemical
potential or self-energies. For the moment we only focus on the
off-diagonal self-energy inhomogeneity and therefore $\check
V_{km}=\delta \Delta_{km} i \hat \sigma_2 i \hat \tau_2$. We begin
by considering a single perturbation of the pairing interaction.
Two different models for the individual pairing perturbation are
represented in the paper. Model (a) describes the region of the
enhanced pair interaction emanating from a site $i$ by Yukawa-type
potential $g_{ij}=g_b+\delta g f_{ij}$, where $g_b$ corresponds to
a background interaction and $f_{ij}=\sum_s
\exp(-r_{ijs}/\lambda)/r_{ijs}$.
$r_{ijs}=\sqrt{r_{ijs}^{(ab)2}+z^2}$ is the distance between the
center of the bond connecting the sites $i$ and $j$ and the source
of the pairing interaction perturbation beyond the plane. Here $z$
is the distance between the $ {\rm CuO}_2 $ plane and the source
of the perturbation and $r_{ijs}^{(ab)}$ denotes the distance in
the plane. Model (b) supposes that the pair interaction is
considerably enhanced on a plaquette and has a long weak tail
beyond this plaquette: $g_{ij}=g_b+u_1$ if the bond $ij$ belongs
to the chosen plaquette and $g_{ij}=g_b+u_2/(r_{0,ij}^2+z^2)$ if
the bond $ij$ is beyond this plaquette. Here, $r_{0,ij}$ is the
distance between the center of the chosen plaquette and the center
of the bond $ij$. By considering a number of different models for
the single perturbation we have checked that the results do not
depend qualitatively on the particular shape of the perturbation
and are only controlled by its effective width and height. The
tight binding model parameters taken to obtain the normal
quasiparticle dispersion are the following: the next-nearest
neighbor hopping $t'=-0.3$ and the chemical potential $\mu=-1$
correspond to the normal state background of model (a) and
$t'=-0.35$ and $\mu=-0.8$ describe the normal background of model
(b). Both sets of the parameters give approximately the same
low-energy dispersion and adjust to qualitatively reproduce
experimentally measured Fermi surface of BSCCO near optimal
doping. However, they result in quite different quasiparticle
dispersions for the energies, which have the absolute value of the
order of superconducting gap, and, in particular, differ by the
energy location of the normal state van Hove singularity
$\varepsilon_{vH}=-\mu + 4 t'$. This fact leads to essential
difference in the shape of the LDOS curves, but does not influence
qualitatively the statistical properties considered here: gap
($\Delta_g$) distribution, the local temperature of vanishing the
gap ($T_p$) distribution and the ratio $2\Delta_g/T_p$. We comment
on the role of the particular set of tight binding model
parameters in more detail below. Making use of the $T$-matrix
approach outlined above we calculate the superconducting order
parameter, local density of states and $T_p$ for the two
inhomogeneous patterns with a single perturbation described by
models (a) and (b), respectively.

\begin{figure}[!tbh]
\begin{minipage}[b]{\linewidth}
   \centerline{\includegraphics[clip=true,width=1.7in]{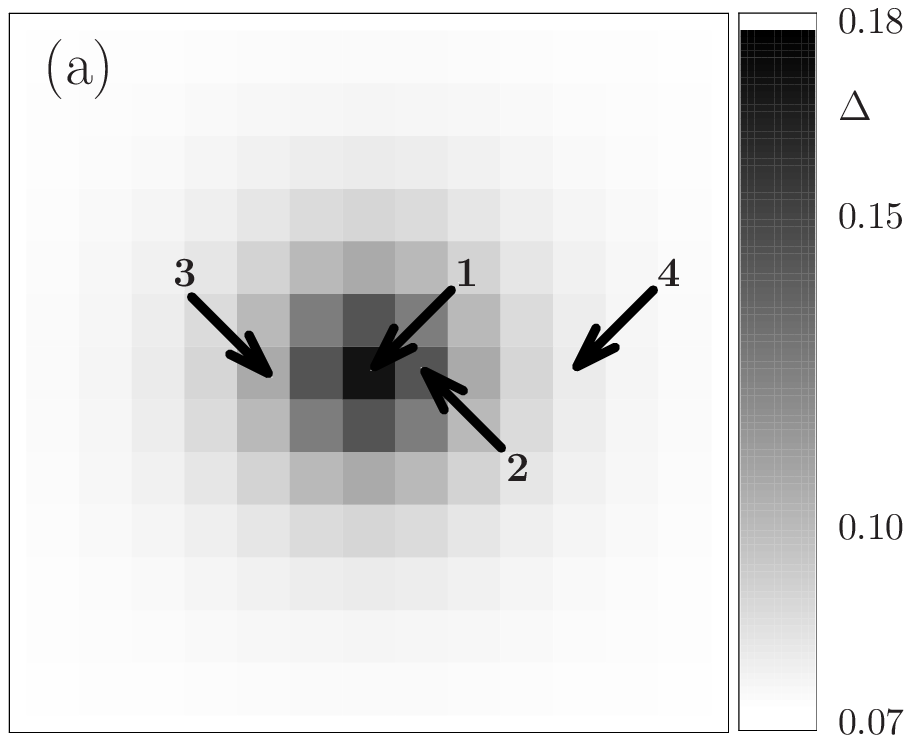}}
  \end{minipage}\hfill
 \begin{minipage}[b]{\linewidth}
   \centerline{\includegraphics[clip=true,width=1.7in]{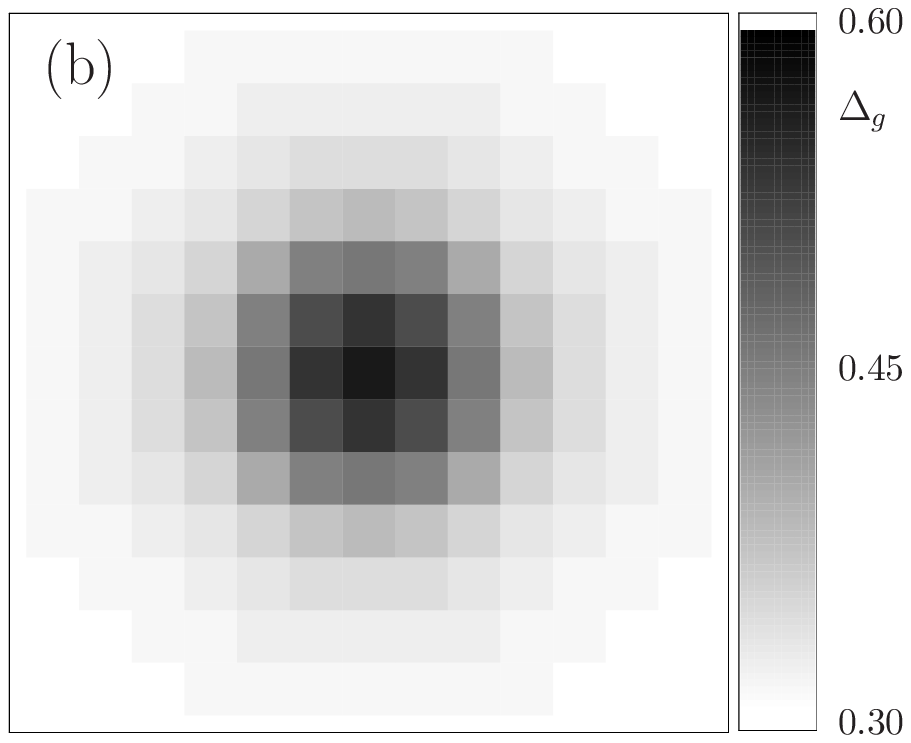}}
  \end{minipage}
  \begin{minipage}[b]{\linewidth}
   \centerline{\includegraphics[clip=true,width=1.7in]{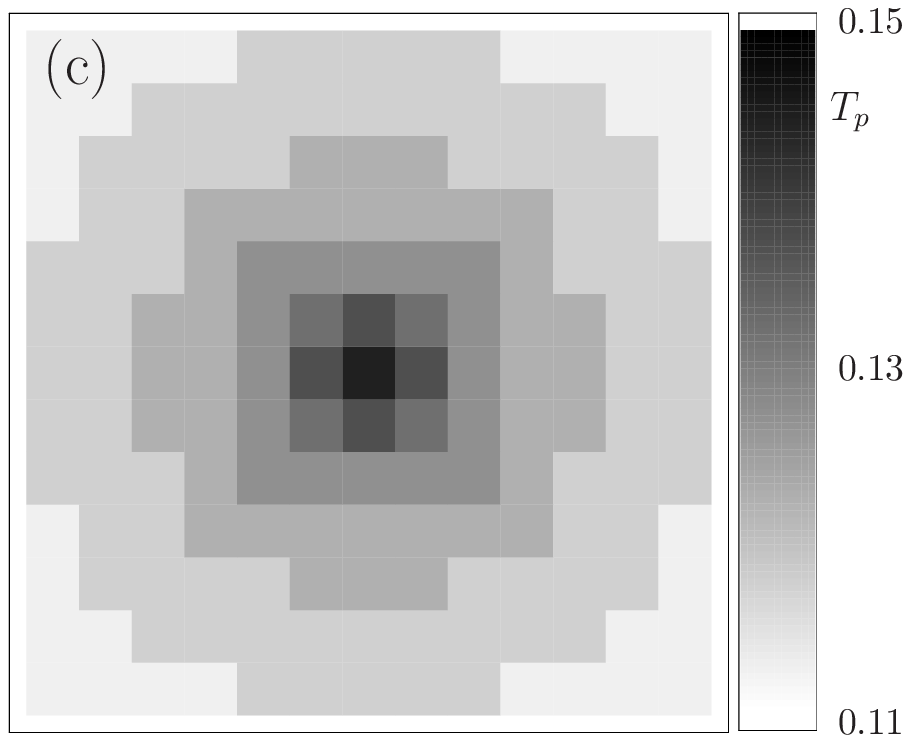}}
  \end{minipage}\hfill
 \begin{minipage}[b]{\linewidth}
   \centerline{\includegraphics[clip=true,width=1.7in]{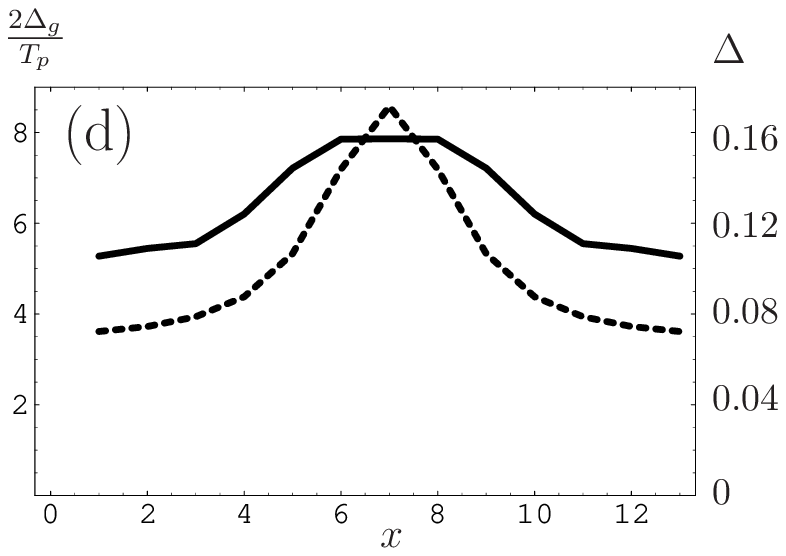}}
  \end{minipage}
  \caption{(a) Low-temperature ($T=0.03$) distribution of the site-averaged
  OP.
  (b) Low-temperature ($T=0.03$) gap map. (c) $T_p-$map. (d) The space profiles
  of the ratio $2 \Delta_g/T_p$ (solid line) and the site-averaged OP (dashed line)
  calculated along the horizontal line drawn though the center of the perturbation
  in panel (a).
  The left (right) vertical axis corresponds to the ratio $2\Delta_g(i)/T_p(i)$
  (superconducting order parameter). All the pictures correspond
  to the single perturbation described by model (a) (see text) with $g_b=0.51$, $\delta g=2.09$,
  $\lambda=1.5$ and $z=1.5$. The parameters $\lambda$ and $z$ are measured
  in units of the lattice constant $a$.} \label{Delta_single_1}
\end{figure}

\begin{figure}[!tbh]
\begin{minipage}[b]{\linewidth}
   \centerline{\includegraphics[clip=true,width=1.7in]{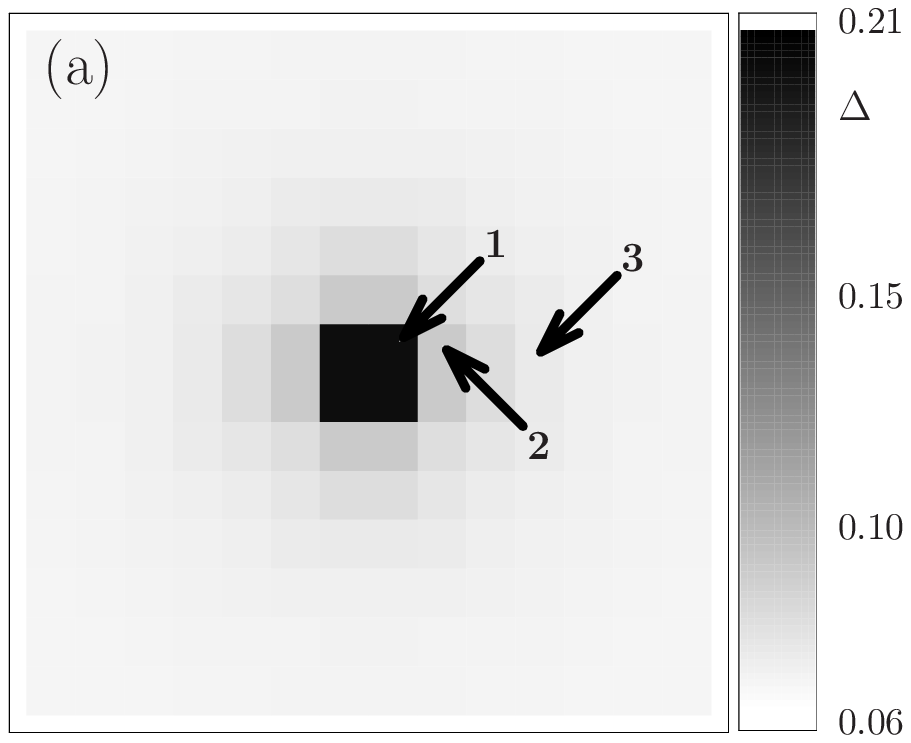}}
  \end{minipage}\hfill
 \begin{minipage}[b]{\linewidth}
   \centerline{\includegraphics[clip=true,width=1.7in]{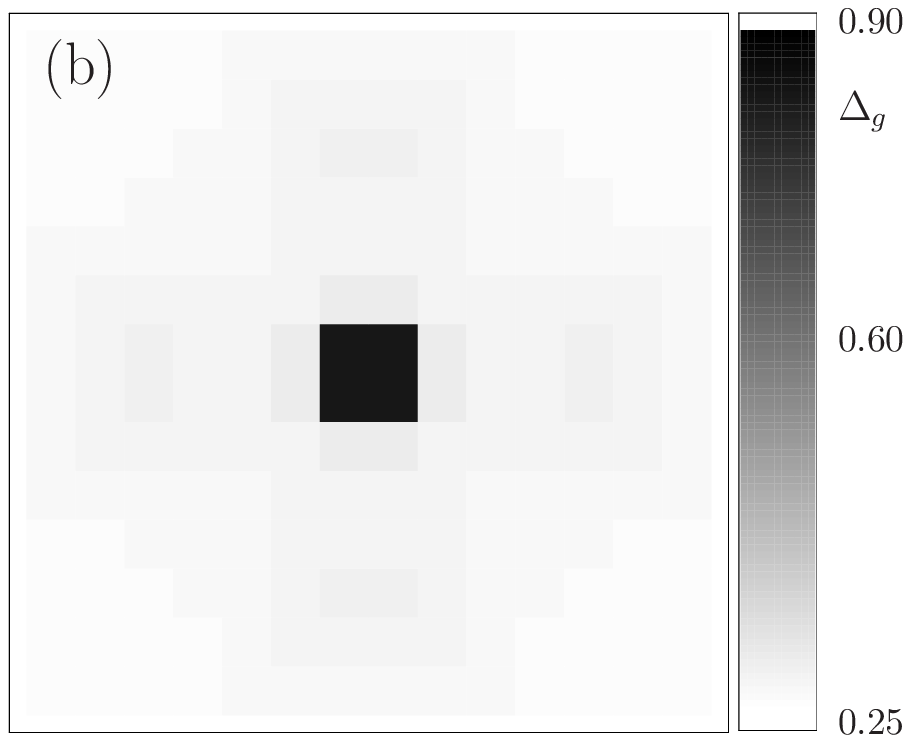}}
  \end{minipage}
  \begin{minipage}[b]{\linewidth}
   \centerline{\includegraphics[clip=true,width=1.7in]{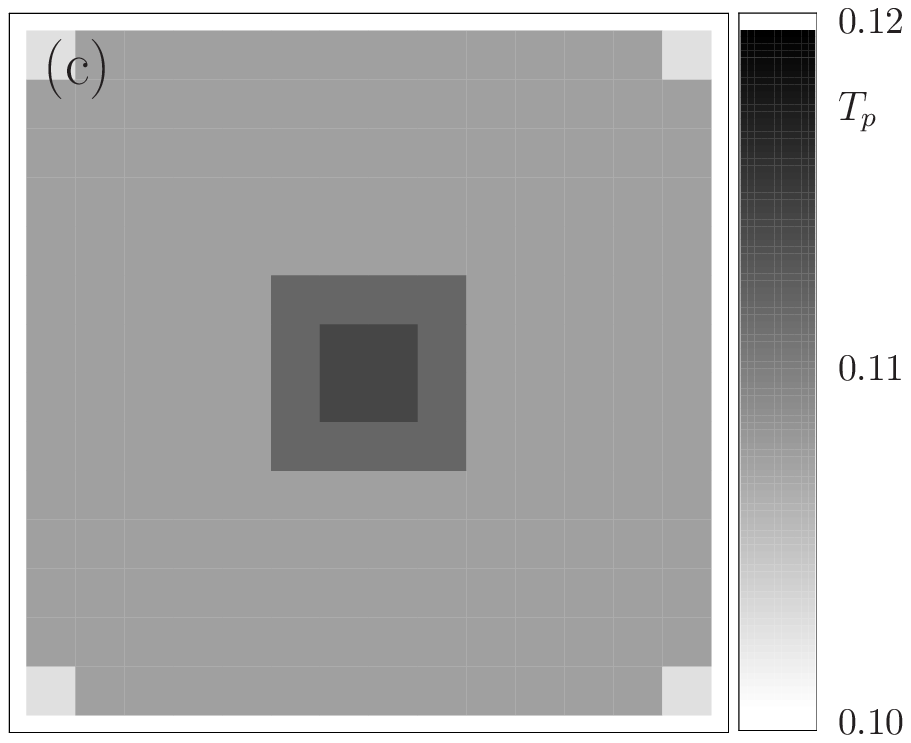}}
  \end{minipage}\hfill
 \begin{minipage}[b]{\linewidth}
   \centerline{\includegraphics[clip=true,width=1.7in]{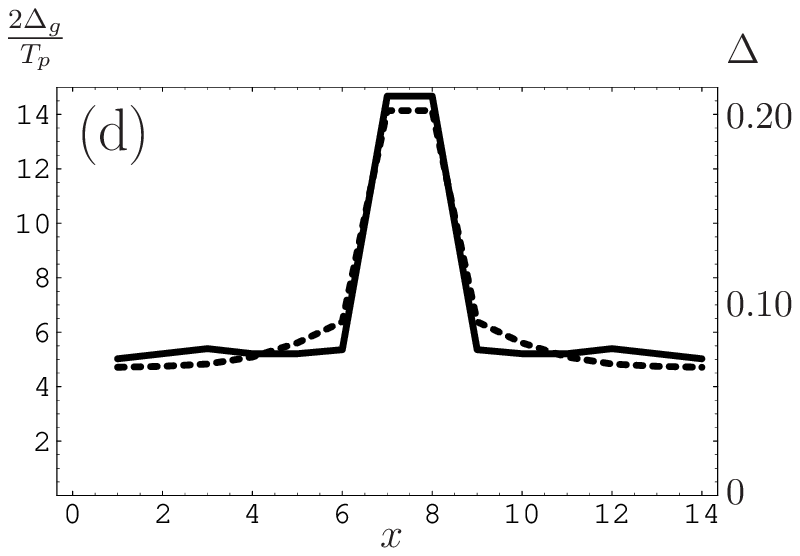}}
  \end{minipage}
  \caption{The panels (a)-(d) represent the same
  quantities as in Fig.~\ref{Delta_single_1}, but correspond
  to the single perturbation described by model (b) (see text) with $g_b=0.59$, $u_1=1.00$,
  $u_2=0.25$ and $z=0.5$. } \label{Delta_single_2}
\end{figure}

The resulting space distributions of low-temperature site-averaged
OP $\Delta_i \equiv (\Delta_{i,i+\hat a}+\Delta_{i,i-\hat
a}+|\Delta_{i, i+\hat b}|+|\Delta_{i,i-\hat b}|)/4$ are presented
in Figs.~\ref{Delta_single_1}(a) and \ref{Delta_single_2}(a) for
models (a) and (b), respectively.

STM technique measures the local differential conductance, and the
thermally smeared LDOS, described by the expression
\begin{equation}
dI/dV \sim \int \limits_{-\infty}^{\infty} d\varepsilon
\frac{df(\varepsilon+V)}{dV}\rho_i(\varepsilon,T)
\label{smeared_DOS} \enspace ,
\end{equation}
can be extracted from these measurements. We set $|e|=1$
throughout the paper. In the above expression
\begin{equation}
\rho_i(\varepsilon,T)=-(1/\pi){\rm Im}G_{ii}^R(\varepsilon,T)
\label{ldos}
\end{equation}
is the local density of states, $f(\varepsilon)$ is Fermi
distribution function and $V$ is voltage applied between the STM
tip and the sample. For the homogeneous situation the distance
between the coherence peaks $2\Delta_g$ in the low-temperature
conductance spectra is directly connected to the low-temperature
OP by the simple relation $\Delta_g=\alpha \Delta_0(T)$, where the
parameter $\alpha$ is slightly model dependent. In our case
$\alpha=4.0$ for model (a) and $\alpha=3.6$ for model (b).

For an inhomogeneous system, where the characteristic size of the
patch $\lesssim \xi$, there is no any direct simple relationship
between the local OP and the local gap $\Delta_g$. One can only
conclude from the numerical calculations \cite{nunner05,fang06}
that for regions, where the OP is enhanced from the background,
the gap gets wider and the peak height is suppressed compared to
the average value. It is worth noting that, unlike the homogeneous
situation, this peak does not always represent the maximal
superconducting gap on the Fermi surface, but in some cases
originates from the spectral weight transfer from the nearby van
Hove singularity due to the Andreev scattering processes.
Otherwise, if the OP in a cluster is less than that one in the
background, the narrow and high Andreev resonant peaks develop in
the cluster region resulting in diminishing of the gap region. For
this reason we investigate not only the OP distribution, but also
the experimentally measurable thermally smeared LDOS, which has a
maximum at $V=\Delta_g$. For models (a) and (b) of the single
perturbation the corresponding low-temperature gap maps are
represented in Figs.~\ref{Delta_single_1}(b) and
\ref{Delta_single_2}(b), respectively.

Experimentally \cite{gomes07} the temperature $T_p(i)$ of the gap
disappearing for the particular location in the sample has been
determined using the criterion $dI/dV(V=0) \ge dI/dV$ (for all $V
\ge 0$). Using the above criterion we calculated the distribution
$T_p(i)$ from the thermally smeared LDOS curves. The corresponding
$T_p$-maps for the single perturbations described by models (a)
and (b) are represented in Fig.~\ref{Delta_single_1}(c) and
\ref{Delta_single_2}(c), respectively. It can be seen from
Figs.~\ref{Delta_single_1} and \ref{Delta_single_2} that while the
space distribution of the superconducting order parameter mainly
follows that one of the pairing interaction strength due to the
local multiplication of the anomalous Green's function in the
self-consistency equation (\ref{self_consistency}) by the coupling
constant, LDOS and $T_p$ are essentially nonlocal (on the atomic
scale) quantities with the characteristic size of order of $\xi_s$
determined by the scale, where the Green's functions change
considerably. This is valid for the both considered models,
however the non-locality is much less pronounced for model (b).
The reason is that the pair interaction perturbation is more
spikier in this case as compared to model (a). Due to its very
small effective width the perturbation (b) cannot contribute
considerably to the Green's function and, consequently, to the
LDOS anywhere in space except for its own plaquette. The reason
for enhancement of the gap at this plaquette is spectral weight
transfer from the nearby Van Hove singularity, which is discussed
in detail later.

The ratios $2\Delta_g(i)/T_p(i)$ calculated along the horizontal
lines drawn through the centers of the perturbations in
Figs.~\ref{Delta_single_1}(a) and \ref{Delta_single_2}(a) for the
model samples considered above are plotted in
Figs.~\ref{Delta_single_1}(d) and \ref{Delta_single_2}(d) together
with the spacial profiles of the superconducting order parameter
along the same cuts. It is seen that this ratio is considerably
enhanced in comparison with the homogeneous situation. For model
(a) the area of the enhancement is larger that the characteristic
size of the superconducting order parameter (and coupling
constant) perturbation. This is due to the above mentioned gap and
$T_p$ nonlocality and leads to the increasing of the ratio over
the entire sample in the situation with many off-diagonal
scatterers in spite of the fact that only smaller part of the
sample is occupied by the enhanced coupling constant regions. On
the contrary, for model (b) the two profiles practically coincide.
Note that the maximal ratio is very high ($\sim 14$) for this
model. The physical reasons for this fact will be given later. The
averaged ratio $\langle 2\Delta_g(i)/T_p(i) \rangle$ calculated
over the region of the size, which approximately corresponds to
the average distance between the centers of enhanced pairing
regions in the situation with many individual perturbations
considered below, is equal to $7.6$ and $8.5$ for models (a) and
(b), respectively.

\begin{figure}[!tbh]
\begin{minipage}[b]{\linewidth}
     \centerline{\includegraphics[clip=true,width=1.7in]{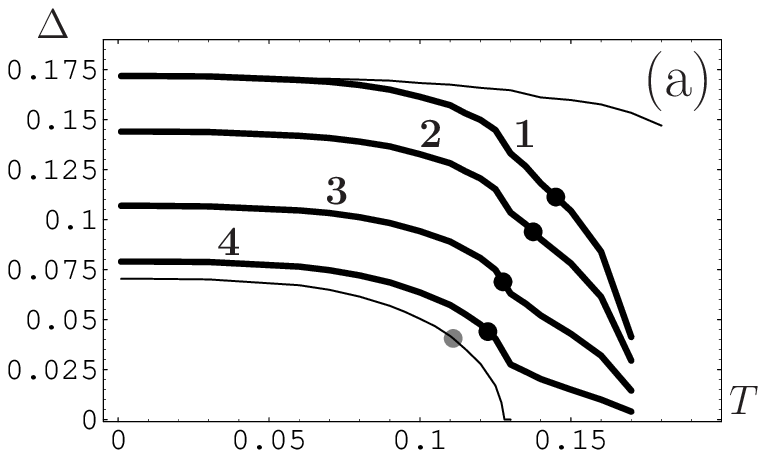}}
     \end{minipage}\hfill
     \begin{minipage}[b]{\linewidth}
   \centerline{\includegraphics[clip=true,width=1.7in]{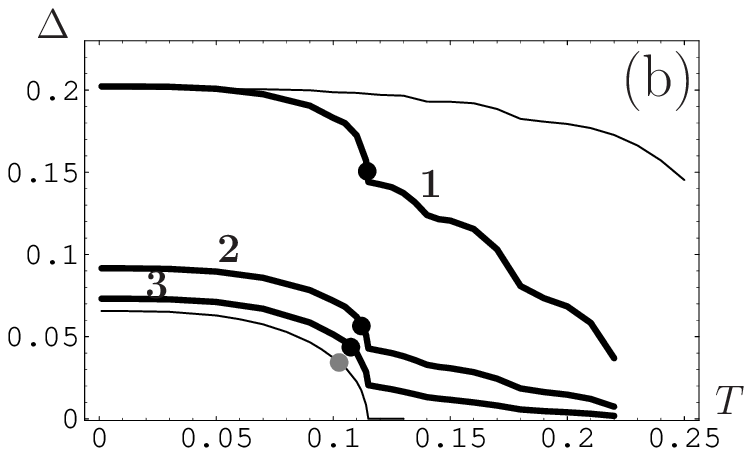}}
  \end{minipage}
        \caption{Dependence of the site-averaged OP
        on temperature for the single perturbation. The thin bottom line corresponds to bulk
        order parameter generated by the background coupling
        constant. The bold solid curves represent the temperature dependence
        of the superconducting order parameter for the sites marked by the
        appropriate numbers in Fig.~\ref{Delta_single_1}(a) and \ref{Delta_single_2}(a). The upper
        thin line is the temperature dependence of the bulk OP
        with the same zero-temperature value as in the center of
        the perturbation (curve marked by $1$). The panels (a) and
        (b) correspond to the models (a) and (b), respectively.}
\label{Delta_T_single}
\end{figure}

There are two main physical reasons for the enhancement of the
ratio $2\Delta_g/T_p$ for the single perturbation considered
above. First of all, small as compared to superconducting
coherence length perturbations cannot maintain superconductivity
by themselves and only do this due to the superconductivity in the
bulk. Although the value of the zero-temperature OP strongly
increases when the height of the perturbation grows, at finite
temperatures where the bulk OP vanishes, the pairing correlations
in the small area go to zero abruptly. To illustrate this the
dependence of the site-averaged OP on temperature is presented in
Fig.\ref{Delta_T_single}. Panels (a) and (b) correspond to models
(a) and (b) described above. The bottom thin line represents the
temperature dependence of the background order parameter generated
by the coupling constant $g_b$. The bold solid curves correspond
to the temperature dependence of the site-averaged order parameter
for the sites marked by the appropriate numbers in
Figs.~\ref{Delta_single_1}(a) and \ref{Delta_single_2}(a),
respectively. Although the superconducting order parameter
vanishes at the critical temperature denoted by $T_x$ over the
entire inhomogeneous sample simultaneously, as it should be in the
mean-field approximation, the pairing correlations in the small
area reduce abruptly near the temperature of the vanishing of the
bulk OP corresponding to the coupling constant $g_b$. Then only
the non BCS-like tails extend up to $T_x$. In other words, the OP
inside the area of the enhanced pair interaction disappear with
temperature more faster than the bulk OP having the same
low-temperature value. To demonstrate this the temperature
dependence of the bulk OP having the same low-temperature value as
in the center of the perturbation (curves marked by $1$) is
depicted by the upper thin lines in Figs. \ref{Delta_T_single}(a)
and (b). Note that for model (b) the abrupt suppression of the OP
near the background critical temperature $T_b$ is seen more
clearly due to smaller effective size of the perturbation.

The temperatures $T_p$ of vanishing the gap for the given space
locations are marked by the filled circles on the corresponding
$\Delta(T)$ curves in Figs. \ref{Delta_T_single}(a) and (b). It is
seen that even for the homogeneous case $T_p$ is slightly lower
than the critical temperature (see the bottom thin line) due to
the thermal smearing of the measured differential conductance
according to the formula (\ref{smeared_DOS}). Consequently, the
bulk ratio $2\Delta_g/T_p$ is a bit higher than the ratio
$2\Delta_g/T_b$. These ratios are dependent slightly on the
particular coupling constant and tight-binding parameters. For
model (a) we consider they have the following values:
$(2\Delta_g/T_b)^{(a)}=4.4$ and $(2\Delta_g/T_p)^{(a)}=5.0$, while
for model (b) $(2\Delta_g/T_b)^{(b)}=4.2$ and
$(2\Delta_g/T_p)^{(b)}=4.7$. For the inhomogeneous situation with
the single perturbation the difference between $T_x$ and $T_p$
($T_p$ is now position dependent) is much more pronounced. This
results from the discussed above non BCS-like behavior of the OP
on temperature: for the long OP tails, taking place for the
temperatures higher than the background critical temperature, the
gap is too small and easily smeared out by high enough
temperature. Although at the center of the perturbation the value
of the superconducting OP is still high enough at the temperatures
of the order of $T_p$, the area occupied by this enhanced OP is
small as compared to $\xi_s^2$. Consequently, the amplitude of the
corresponding peak in the LDOS is low and again can be easy washed
out by temperature. These reasons work more efficiently for more
spikier perturbations. This is clearly seen in
Fig.\ref{Delta_T_single}, where the difference between $T_p$ and
$T_x$ is more pronounced for model (b).

When the size of the pairing interaction perturbation increases,
$T_x$ grows considerably, while the zero-temperature value of the
superconducting OP in the perturbation region rises only slightly.
Naturally, when the size of the enhanced pairing area becomes of
the order of a few superconducting coherence lengths, $T_x \to
T_b$ corresponding to the coupling constant $g_b+\delta g$ and the
bulk value of $2\Delta_g/T_p$ should be restored at the center of
the cluster. For $s-wave$ pairing case and cluster width
comparable to or larger than $\xi_s$ the reduction of the critical
temperature of the cluster having the pairing potential $g_1$ (in
comparison to the bulk value corresponding to $g_1$) due to the
proximity to the background with the pairing potential $g_0<g_1$
has been studied in Ref.\cite{ovchinnikov01} in the framework of
quasiclassical Usadel equations. As it is seen from the above
picture, the OP inhomogeneity of the size comparable to or larger
than $\xi_s$ cannot give rise to large enough ratios
$2\Delta_g/T_p$ and more spikier atomic-scale inhomogeneities are
more favorable.

\begin{figure}[!tbh]
\begin{minipage}[b]{\linewidth}
     \centerline{\includegraphics[clip=true,width=1.7in]{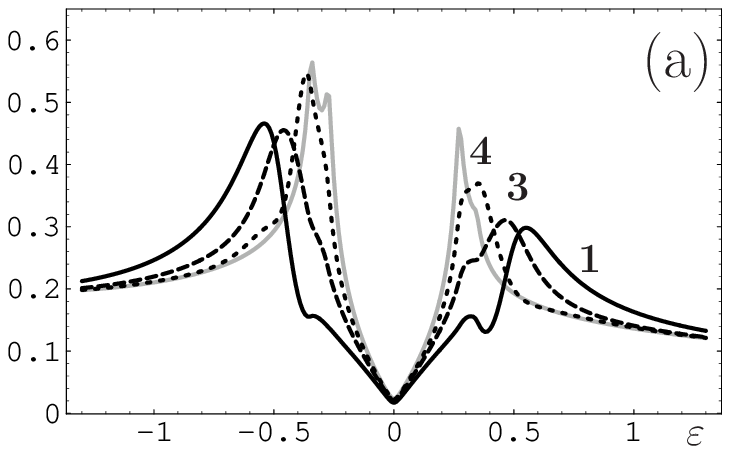}}
     \end{minipage}\hfill
     \begin{minipage}[b]{\linewidth}
   \centerline{\includegraphics[clip=true,width=1.7in]{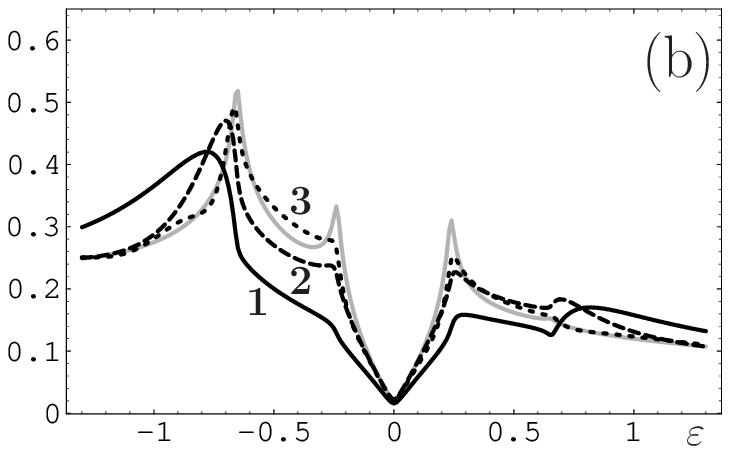}}
  \end{minipage}
        \caption{Low-temperature ($T=0.03$) LDOS curves taken for
        a number of sites from the center of the perturbation
        (black solid curve) to the background (gray line).
        The curves correspond to the sites marked by the
        appropriate numbers in
        Fig.~\ref{Delta_single_1}(a) and
        Fig.~\ref{Delta_single_2}(a). Panels (a) and (b)
        correspond to models (a) and (b), respectively.}
        \label{ldos_single}
\end{figure}

The second reason for the ratio $2\Delta_g/T_p$ to get larger is
the following. As it was already mentioned above, in the
inhomogeneous situation with an atomic-scale single perturbation
of the large enough strength the main peak in LDOS does not
represent the maximal local superconducting gap on the Fermi
surface, but originates from the spectral weight transfer from the
nearby van Hove singularity due to the Andreev scattering
processes. For the tight-binding parameters we consider the van
Hove singularity energy in the homogeneous system is approximately
described by the formula $\varepsilon_{vH}=-\sqrt{(-\mu+4
t')^2+\Delta_{max}^2}$, where $\Delta_{max}$ is an effective
maximal superconducting gap on the Fermi surface. The processes of
Andreev scattering from the inhomogeneity result in appearing of
the symmetrical with respect to Fermi energy peak in LDOS. The
examples of low-temperature LDOS curves exhibiting such a feature
are shown in Fig.~\ref{ldos_single}. Panels (a) and (b) correspond
to models (a) and (b), respectively. It is seen that for model
(a), where the normal state van Hove singularity is very close to
the Fermi energy (and, consequently, the van Hove singularity peak
in the superconducting state is close to the superconducting
coherence peak), the above mentioned transfer of the spectral
weight leads to small enough increase of the gap $\Delta_g$. The
shape of the resulting peak is close to the shape of a
superconducting coherence peak because of partial overlapping of
the van Hove singularity and superconducting coherence peak in the
bulk. At the same time for model (b), where the van Hove
singularity is more distinct from the superconducting coherence
peak in the bulk, the clearly seen spectral weight transfer
results in considerable increase of the gap $\Delta_g$. However,
this mechanism of a gap enhancement is very local, as it is seen
in Fig.~\ref{ldos_single}(b). This is the reason for appearing of
very high gap just at the plaquette occupied by the perturbation
in Fig.~\ref{Delta_single_2}(b). However, the shape of the
resulting peak is strongly distorted. It is worth to note that the
LDOS curves represented here are related to the single
perturbation. For the many individual perturbations discussed
below the curves have more resemblance to the experimentally
measured LDOS spectra.

\subsection{many OP scatterers}
\label{many_phys}

\begin{figure}[!tbh]
\begin{minipage}[b]{\linewidth}
   \centerline{\includegraphics[clip=true,width=1.7in]{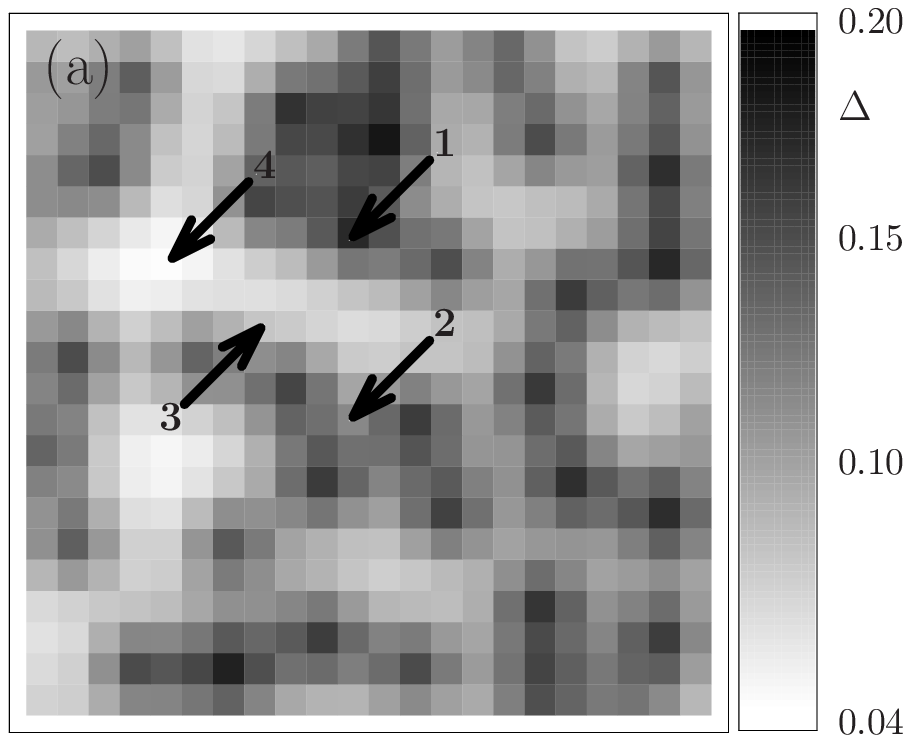}}
  \end{minipage}\hfill
 \begin{minipage}[b]{\linewidth}
   \centerline{\includegraphics[clip=true,width=1.7in]{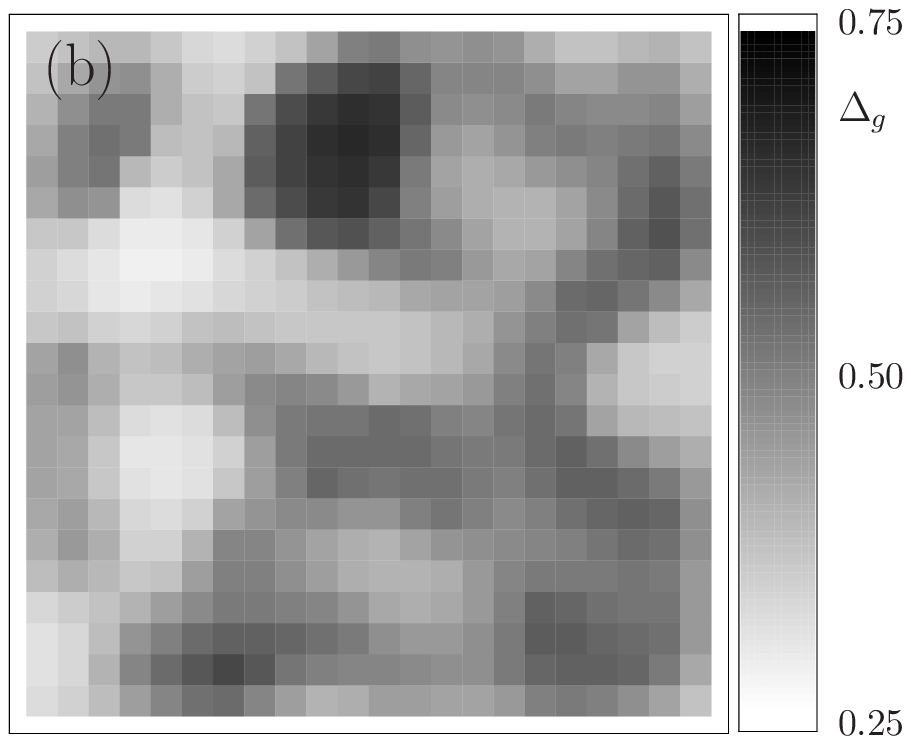}}
  \end{minipage}
  \begin{minipage}[b]{\linewidth}
   \centerline{\includegraphics[clip=true,width=1.7in]{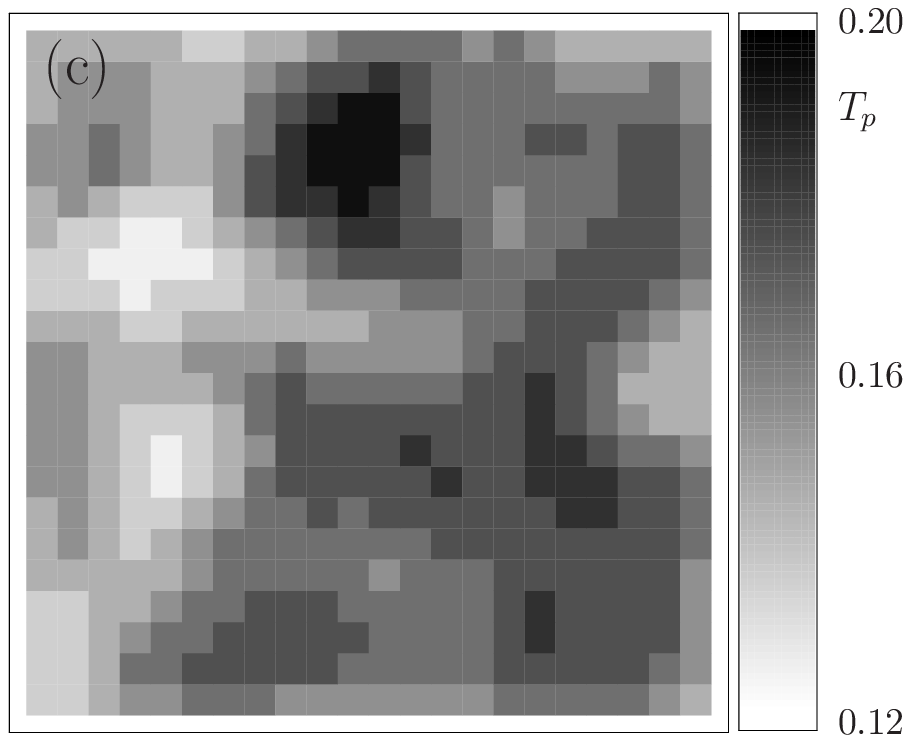}}
  \end{minipage}\hfill
 \begin{minipage}[b]{\linewidth}
   \centerline{\includegraphics[clip=true,width=1.7in]{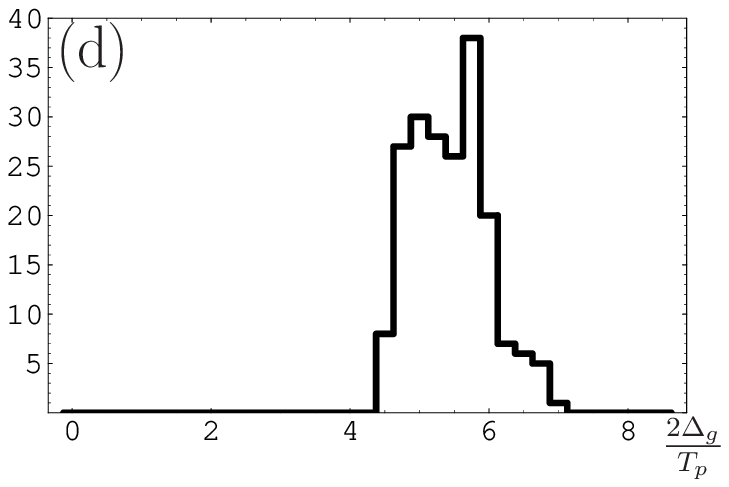}}
  \end{minipage}
  \caption{(a) Low-temperature ($T=0.05$) distribution of the site-averaged
  OP for model (A). (b) Low-temperature gap map. (c) $T_p-$map.
  (d) Histogram of $2\Delta_g(i)/T_p(i)$ values. The parameters describing
  a perturbation are the following: $g_b=0.05$, $\delta g=2.72$, $\lambda=z=1.5$.} \label{Delta_many_1}
\end{figure}

\begin{figure}[!tbh]
\begin{minipage}[b]{\linewidth}
   \centerline{\includegraphics[clip=true,width=1.7in]{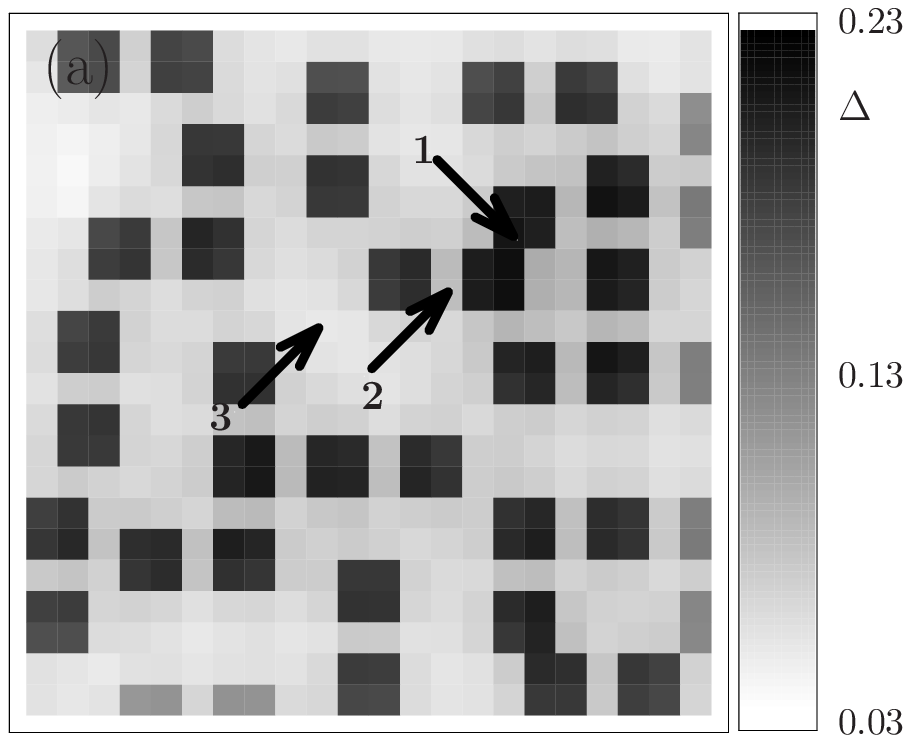}}
  \end{minipage}\hfill
 \begin{minipage}[b]{\linewidth}
   \centerline{\includegraphics[clip=true,width=1.7in]{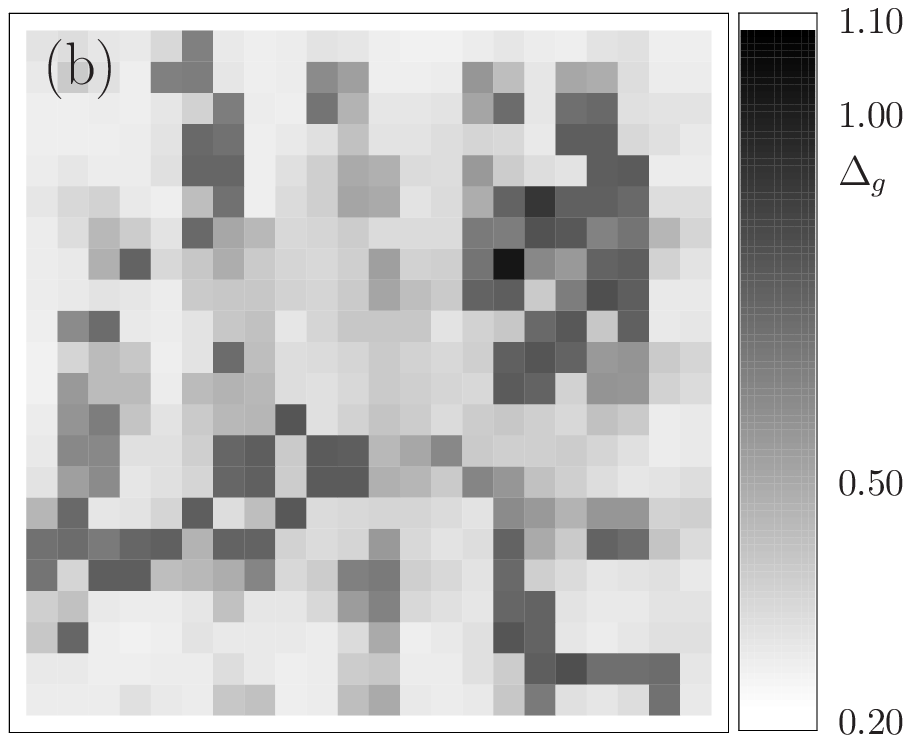}}
  \end{minipage}
  \begin{minipage}[b]{\linewidth}
   \centerline{\includegraphics[clip=true,width=1.7in]{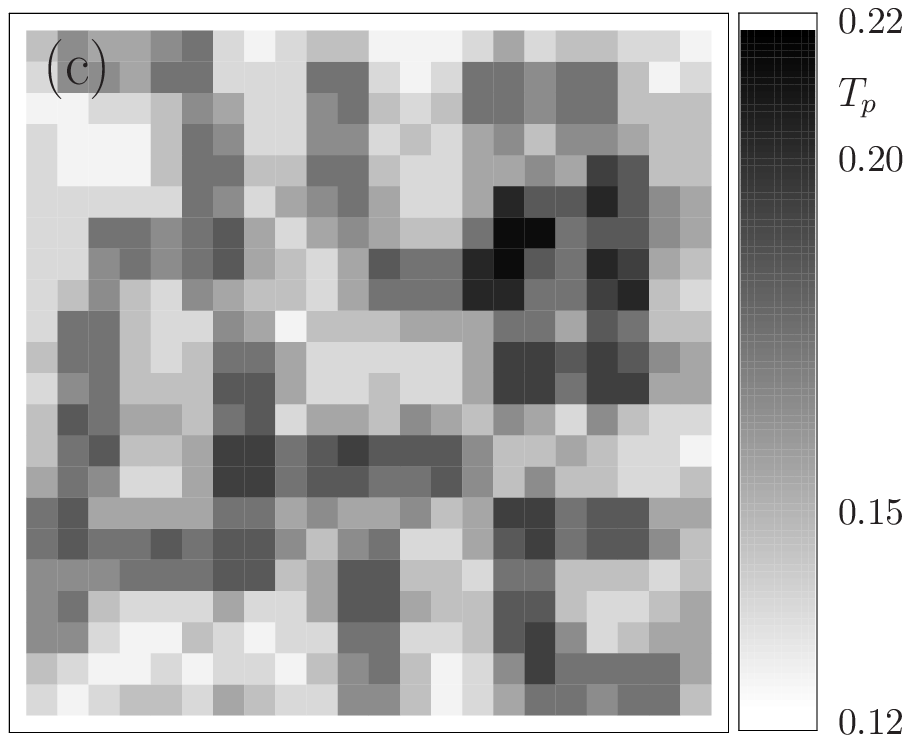}}
  \end{minipage}\hfill
 \begin{minipage}[b]{\linewidth}
   \centerline{\includegraphics[clip=true,width=1.7in]{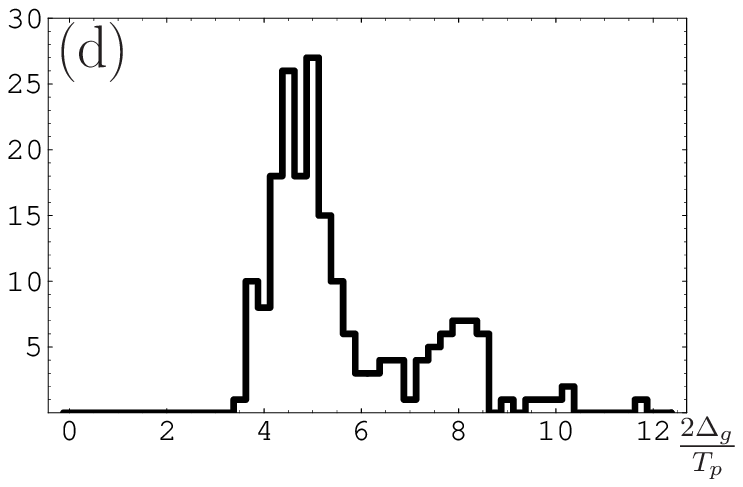}}
  \end{minipage}
  \caption{(a) Low-temperature ($T=0.03$) distribution of the site-averaged
  OP for model (B). (b) Low-temperature gap map. (c) $T_p-$map.
  (d) Histogram of $2\Delta_g(i)/T_p(i)$ values. The individual perturbation
  and the background are described by the following parameters:
  $g_b=0.25$, $u_1=1.16$, $u_2=0.25$ and $z=0.5$.} \label{Delta_many_2}
\end{figure}

Now we turn to discussion of more realistic situation, when many
pair interaction scatterers are present in the sample. We have
considered a $22 \times 22$ sites square as a perturbation
described by the T-matrix. We again represent here two models. In
the both models individual perturbations are randomly distributed
in the square with the concentration $n=0.07$ (here concentration
means the number of individual scatterers divided by the total
number of sites in the square). In model (A) the individual
perturbations have the same shape as in the single perturbation
model (a). Model (B) corresponds to the individual perturbations
described by model (b). The tight-binding parameters describing
the normal state quasiparticle dispersion coincide for models (a)
and (A) so as for models (b) and (B), respectively. The numerical
values of the parameters, characterizing the individual
scatterers, are taken to give the same low-temperature values of
superconducting OP in the center of the perturbation and in the
background as for the corresponding single perturbation model (see
captions to Figs.~\ref{Delta_many_1} and \ref{Delta_many_2} for
the particular values).

Since we consider the finite-size square as a perturbation
described by $T$-matrix, the finite-size effects will affect the
results. However the appropriate choice of the homogeneous
superconducting OP $\Delta^0$ outside the square helps us to
minimize the influence of the square boundary. By considering a
larger $30 \times 30$ sites square we have checked that if the
bulk OP outside the square is taken to be equal to the average
background value of the OP in the square, the error in calculating
OP, gap and $T_p$ becomes less than $5\%$ at the distance $\sim 4$
sites from the boundary and lessens further towards the center of
the square. For this reason the $14 \times 14$ central region of
the square is taken into account in calculating the statistical
properties of the system.

The resulting spatial distributions of low-temperature
site-averaged OP, low-temperature gap-maps and $T_p$ maps are
demonstrated in Figs.~\ref{Delta_many_1} and \ref{Delta_many_2}
for models (A) and (B), respectively. It is worth to note that the
temperature evolution of OP spatial distribution was already
discussed in Ref.~\onlinecite{andersen06}, so we do not focus here
on this aspect. The spatial distributions of low-temperature OP
are represented in Figs.~\ref{Delta_many_1}(a) and
\ref{Delta_many_2}(a). As it was described earlier for the single
perturbation, the spatial profile of the OP inhomogeneity mainly
follows that one of the coupling constant, that is the
characteristic scale of the corresponding inhomogeneity is $\sim
2$ atomic sites. However, due to discussed above fact that the
deviation of the Green's function from the bulk value is spread
over a wider space region than the superconducting OP, the gap and
$T_p$ inhomogeneities are more smooth and the characteristic size
of the patches in the gap and $T_p$ maps is of the order of $4-5$
atomic sites for model (A) and $\sim 3$ atomic sites for model
(B). For model (A) the size of a patch is roughly in accordance
with the superconducting coherence length $\xi_s$, while for
model (B) it is somewhat smaller due to very spiky character of
the perturbation.

The probability distributions to find the particular value for
$2\Delta_g(i)/T_p(i)$ in our model samples are plotted in
Figs.~\ref{Delta_many_1}(d) and \ref{Delta_many_2}(d). It is seen
that the distribution for model (A) is quite narrow. The reason is
that the spatial profile of the ratio $2\Delta_g/T_p$ is broadened
and flattened as compared to OP spatial profile, as it was
demonstrated in Fig.~\ref{Delta_single_1}(d) and discussed in the
context of the single perturbation. For model (B) the distribution
is wider and exhibits a long tail of very high ratios for a small
part of sites. The reason for this fact was discussed in the
context of the single perturbation.

The average values of the ratio $2\Delta_g(i)/T_p(i)$ for models
(A) and (B) are $5.4$ and $5.6$, respectively. These values are
considerably less than the ones for the appropriate single
perturbations, described by models (a) and (b) (as it was
discussed earlier, for the single perturbation the ratio $\langle
2\Delta_g(i)/T_p(i) \rangle$ averaged over the region $\sim l^2$,
where $l$ corresponds to the average distance between the centers
of enhanced pairing regions for the appropriate many scatterers
model, equals to $7.6$ and $8.5$ for models (a) and (b),
respectively). Although the mean-field ratios for many scatterers
models exceed the corresponding bulk ratios
$2\Delta_g^{A,av}/T_p^{A,av}=5.2$ and
$2\Delta_g^{B,av}/T_p^{B,av}=4.8$, this increase is quite small as
compared to the single perturbation, especially for model (A).
Here the homogeneous ratios $2\Delta_g^{A(B),av}/T_p^{A(B),av}$
are calculated for the coupling constants, which give the same
zero-temperature OP, as the average zero-temperature OP in
samples (A) and (B). The reason for such a crucial reduction of
the ratio $2\Delta_g/T_p$ in the presence of many OP scatterers is
the proximity effect. The individual perturbations essentially
interact one with another for the considered concentrations when
the averaged distance between them is less then the
superconducting coherence length. In the framework of the
mean-field treatment the phase of the OP is the same over the
entire sample. For this reason the individual perturbations raise
the pairing correlations in the background and maintain their own
OP even for high enough temperatures in comparison with the
background critical temperature. This is quite different from the
picture of the single perturbation, where the pairing correlations
sharply weaken for the temperatures higher than $T_b$.

\begin{figure}[!tbh]
\begin{minipage}[b]{\linewidth}
     \centerline{\includegraphics[clip=true,width=1.7in]{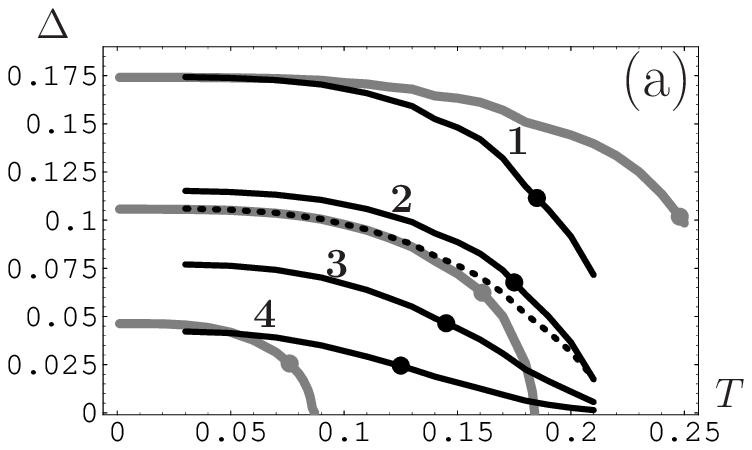}}
     \end{minipage}\hfill
     \begin{minipage}[b]{\linewidth}
   \centerline{\includegraphics[clip=true,width=1.7in]{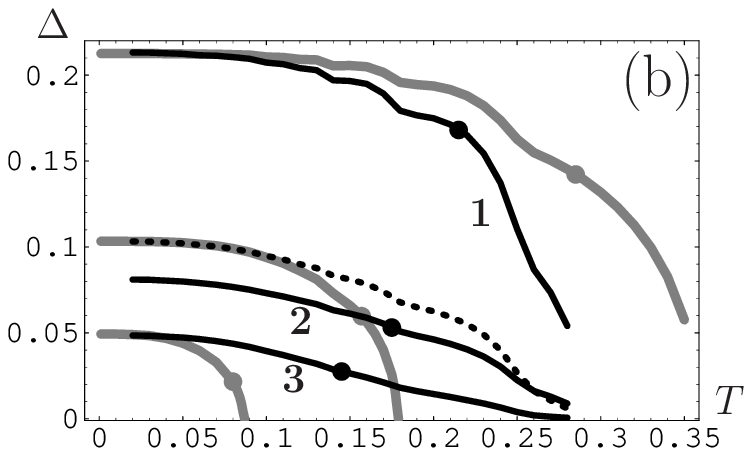}}
  \end{minipage}
        \caption{The dependence of the site-averaged OP on temperature for the
sites marked by the appropriate numbers in
Figs.~\ref{Delta_many_1}(a) and \ref{Delta_many_2}(a) (black solid
lines). The gray lines are the temperature dependence of the bulk
OP having the same zero-temperature value as in the center of the
chosen perturbation (upper gray curve), at the chosen site
belonging to the background (bottom gray curve) and as the
averaged OP (middle gray curve). The dotted line represents the
temperature dependence of the OP, averaged over the entire sample.
Panels (a) and (b) correspond to models (A) and (B),
respectively.} \label{Delta_T_many}
\end{figure}

The discussed behavior of the superconducting OP is depicted in
Fig.~\ref{Delta_T_many}. Panels (a) and (b) correspond to  models
(A) and (B), respectively. The black solid lines represent the
dependence of the site-averaged OP on temperature for the sites
marked by the appropriate numbers in Figs.~\ref{Delta_many_1}(a)
and \ref{Delta_many_2}(a). They correspond to the characteristic
OP behavior along the line from the center of the perturbation to
the background (from top to bottom). The gray lines illustrate the
temperature dependence of the bulk OP having the same
zero-temperature value as in the center of the chosen perturbation
(upper gray curve), at the chosen site belonging to the background
(bottom gray curve) and as the averaged OP (middle gray curve).
The dashed line represents the temperature dependence of the OP,
averaged over the entire sample. Although the curves still exhibit
non-BCS shape, there is no sharp suppression of the
superconducting correlations in the vicinity of the background
critical temperature unlike the single perturbation case. Now,
roughly speaking, it is the critical temperature of the bulk OP,
having the same low-temperature value as the averaged OP (middle
gray curve), that plays part of the background critical
temperature for the single perturbation: if the OP at a given
location is less than the averaged value, the superconducting
correlations for a finite temperature are higher than the bulk
ones corresponding to the same zero-temperature OP, and only for
the locations, where the OP exceeds the average value, the
finite-temperature superconducting correlations are suppressed as
compared to the bulk behavior. This leads to the fact that the
values of $T_p(i)$ are higher than for the single perturbation
case, what, in turn, results in considerable decrease of the ratio
$2\Delta_g/T_p$. It is worth noting that in our mean-field
treatment we neglect OP phase. However, it is physically
reasonable that the phase of the superconducting order parameter
should fluctuate from one region of enhanced pairing amplitude to
another in such an inhomogeneous situation, as it was already
mentioned in the introduction. We discuss the influence of the
thermal phase fluctuations on the above results in the next
section.

\begin{figure}[!tbh]
\begin{minipage}[b]{.5\linewidth}
   \centerline{\includegraphics[clip=true,width=1.7in]{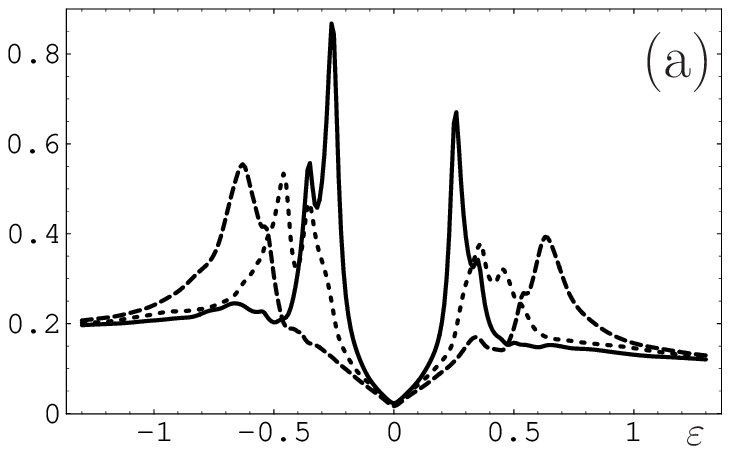}}
  \end{minipage}\hfill
 \begin{minipage}[b]{.5\linewidth}
   \centerline{\includegraphics[clip=true,width=1.7in]{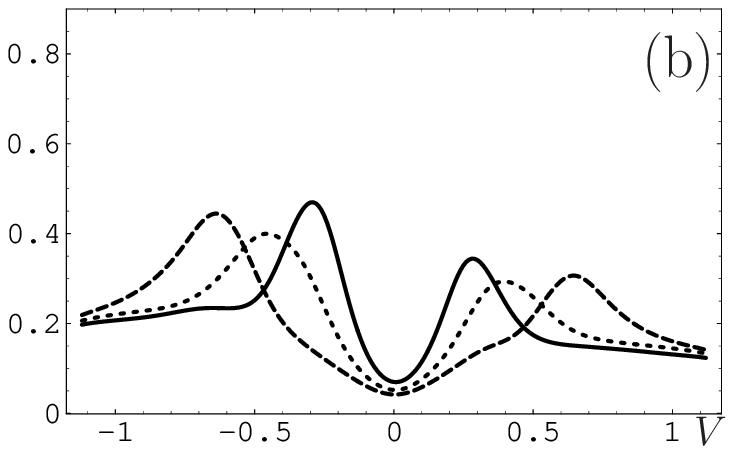}}
  \end{minipage}\hfill
 \begin{minipage}[b]{.5\linewidth}
   \centerline{\includegraphics[clip=true,width=1.7in]{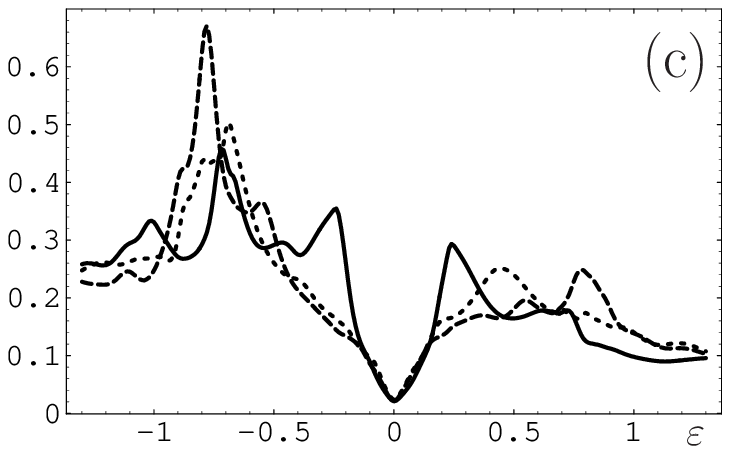}}
  \end{minipage}\hfill
 \begin{minipage}[b]{.5\linewidth}
   \centerline{\includegraphics[clip=true,width=1.7in]{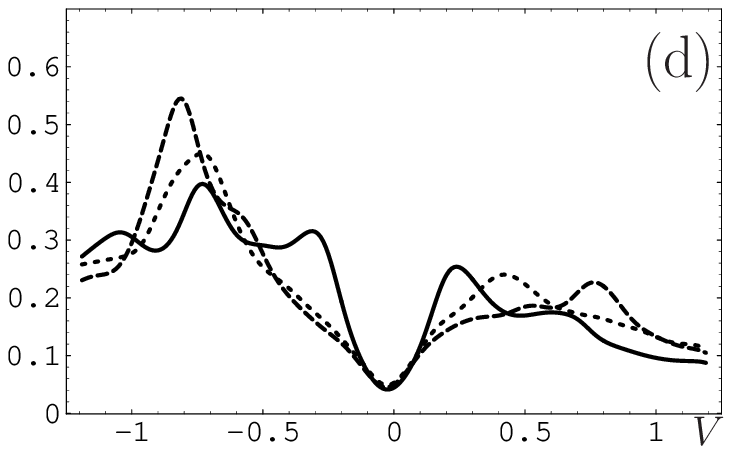}}
  \end{minipage}\hfill
\begin{minipage}[b]{.5\linewidth}
   \centerline{\includegraphics[clip=true,width=1.7in]{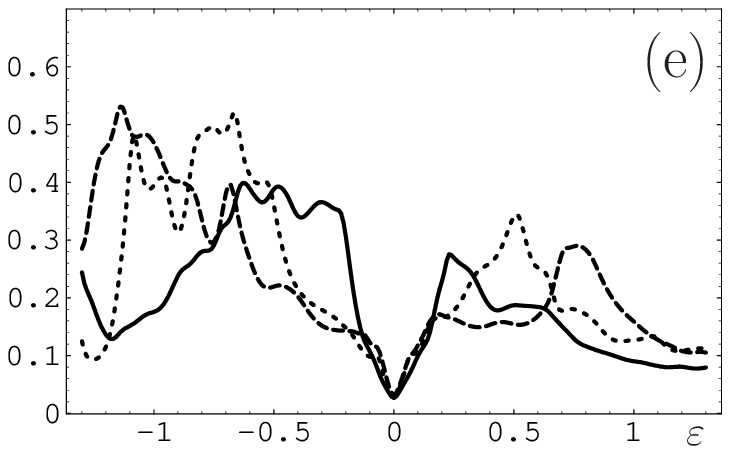}}
  \end{minipage}\hfill
 \begin{minipage}[b]{.5\linewidth}
   \centerline{\includegraphics[clip=true,width=1.7in]{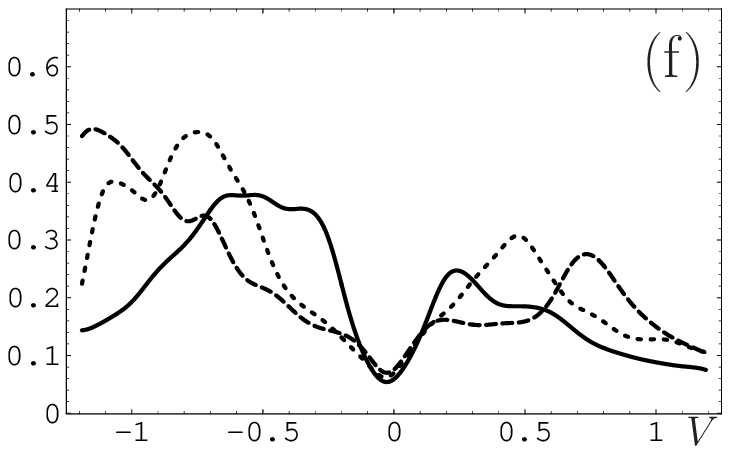}}
  \end{minipage}
    \caption{LDOS (left column) and thermally smeared LDOS (right
    column).
    For all the figures the black solid curve represents LDOS at a given site
    typical for the background, dashed line corresponds
    to the center of a perturbation and LDOS somewhere between
    them is plotted by the dotted line. Upper row represents the results
    for model (A), thermally smeared LDOS is calculated at $T=0.05$.
    Middle row demonstrates the results for model (B), while
    bottom row corresponds to model (B'). Thermally
    smeared LDOS for models (B) and (B') is obtained at $T=0.03$.} \label{ldos_many}
\end{figure}

As it was mentioned above, the difference between the averaged
ratio $\langle 2\Delta_g(i)/T_p(i) \rangle$ for model (B) and the
corresponding homogeneous value is higher than between the same
quantities in model (A). In addition, the probability distribution
to find the particular value of the ratio has a long tail of very
high values for this model. The ratio is greatly enhanced for a
small part of the sites due to the fact that the mechanism of the
spectral weight transfer from the van Hove singularity makes the
gap considerably larger for this model, where the van Hove
singularity and the bulk coherence peak are quite distinct, but it
only works in the close vicinity of an individual perturbation. To
illustrate this the LDOS and thermally smeared LDOS calculated
according to the expression (\ref{smeared_DOS}) at some typical
sites are shown in Figs.~\ref{ldos_many}(a),(b) for model (A) and
Figs.~\ref{ldos_many}(c),(d) for model (B), respectively. The left
column represents the bare LDOS, while the right one corresponds
to the thermally-smeared quantity. Please note that the
temperatures we use to calculate the thermally smeared LDOS are
deeply in the superconducting state, where the superconducting OP
practically does not differ from its zero-temperature value.

It is seen from the figures that for the most part of sites the
bare LDOS exhibits two peaks (and also a number of kinks). The
peak at smaller energies represents a superconducting coherence
peak corresponding to an effective OP. This effective OP neither
coincides with the background value, as it was for the single
perturbation, nor represents the local OP value. Apparently, its
value is close to the average over some area for the most part of
the samples, however this will be discussed in more detail
elsewhere \cite{bobkovy}. The second peak at higher energies is
usually weaker (except for the small part of sites in the sample
(B)) and associated with the spectral weight transfer from the van
Hove singularity due to Andreev scattering. For the both models
the spectral weight transfer is clearly seen in the bare LDOS,
however the two peaks overlap for model (A) due to the fact that
the normal state van Hove singularity is quite close to the Fermi
surface in this case. As a result the effective superconducting
coherence peak and the one transferred from the van Hove
singularity merge in the thermally smeared LDOS even for a low
temperature, as it is seen in Fig.\ref{ldos_many}(b). The
resulting peak resembles a superconducting coherence peak.

For model (B) the effective superconducting coherence peak and the
peak transferred from van Hove singularity are more distinct. In
this case the effective coherence peak is higher than the
transferred one for a number of sites. However, they are of the
same order for the most part of sites. As a result the gap
position is located somewhere between them in the low-temperature
thermally smeared LDOS. The corresponding curves do not exhibit a
pronounced peak and resemble the $dI/dV$ behavior observed in
underdoped samples in contrast to model (A) LDOS, which is more
similar to optimally and overdoped samples. Finally, the
transferred peak wins for a small number of sites, typically at
centers of the individual perturbations. This leads to the sharp
increase of the gap for these spatial locations and, consequently,
to the existence of the long tails in the $2\Delta_g/T_p$
distribution.

\subsection{effect of weak potential and hopping element inhomogeneities}
\label{add_dis}

It is reasonable to assume that the possible causes of the pairing
interaction inhomogeneities (for example, dopant atoms) most
probably also give rise to a disorder of the normal state
parameters. Therefore, let us turn to the discussion of the
additional effect of inhomogeneities of tight binding parameters
such as chemical potential and hopping matrix elements on the
measured by STM properties. Here we only focus on a weak
inhomogeneity of the chemical potential $|\delta \mu| \lesssim t$.
Strong potential scatterers characterizing by $|\delta \mu| \gg t$
are well-known to give rise to quasiparticle resonant bound states
in the close vicinity of the potential impurity \cite{balatsky04}.
In real BSCCO system only very small part of in plane native
defects exhibits near zero bias resonances, so they should not
influence considerably the statistical properties studied here,
such as the $2\Delta_g/T_p$ distribution.

To illustrate the influence of this additional disorder we have
chosen our model (B), where the nearest neighbor hopping element
is taken to be $t+\delta t$, the next-nearest neighbor equals to
$t'+\delta t'$ and the chemical potential is $\mu+\delta \mu$ for
the plaquettes corresponding to the enhanced pairing interaction.
Here we take $\delta t=0.50$, $\delta t'=0.15$ and $\delta
\mu=-0.25$. This model is referred to as (B').

The LDOS and low-temperature thermally smeared LDOS for model (B')
are shown in Figs.~\ref{ldos_many}(e) and (f), respectively. As it
is seen in the figures the low-energy part of the curves is
practically not affected by the potential and hopping matrix
element disorder. Only the high energy part and, especially the
region of the van Hove singularity is distorted by this type of
scatterers. The van Hove singularity become less pronounced under
the influence of the additional disorder: more widened and reduced
in height. This is natural because in the homogeneous situation
the change of tight binding parameters strongly shifts the energy
location of the van Hove singularity. It is worth to note that for
the models, where the normal state van Hove singularity is more
close to the Fermi energy (similar to model (A) we consider), the
effect of the diagonal scatterers on the LDOS is even weaker,
because the van Hove singularity is absorbed by the
superconducting coherence peak and the resulting peak is not
qualitatively sensitive to the discussed types of disorder, as it
was shown in Ref.~\onlinecite{nunner05} for the case of weak
potential disorder.

\begin{figure}[!tbh]
\begin{minipage}[b]{\linewidth}
   \centerline{\includegraphics[clip=true,width=1.7in]{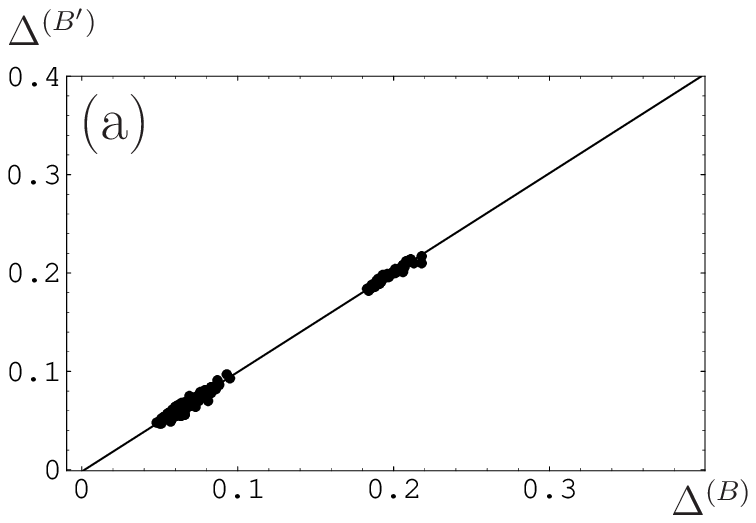}}
  \end{minipage}\hfill
 \begin{minipage}[b]{\linewidth}
   \centerline{\includegraphics[clip=true,width=1.7in]{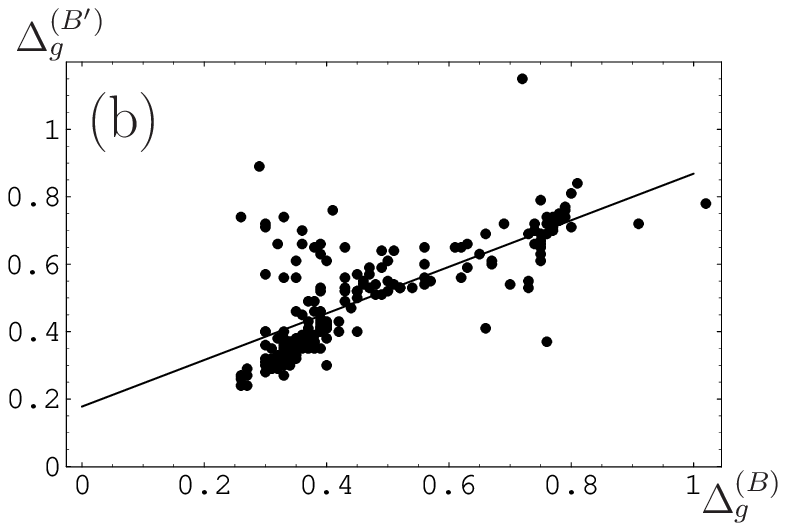}}
  \end{minipage}\hfill
 \begin{minipage}[b]{\linewidth}
   \centerline{\includegraphics[clip=true,width=1.7in]{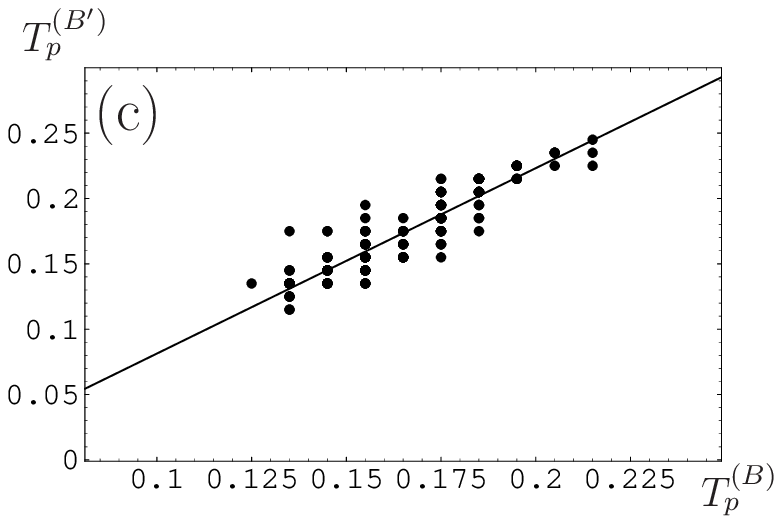}}
  \end{minipage}\hfill
 \begin{minipage}[b]{\linewidth}
   \centerline{\includegraphics[clip=true,width=1.7in]{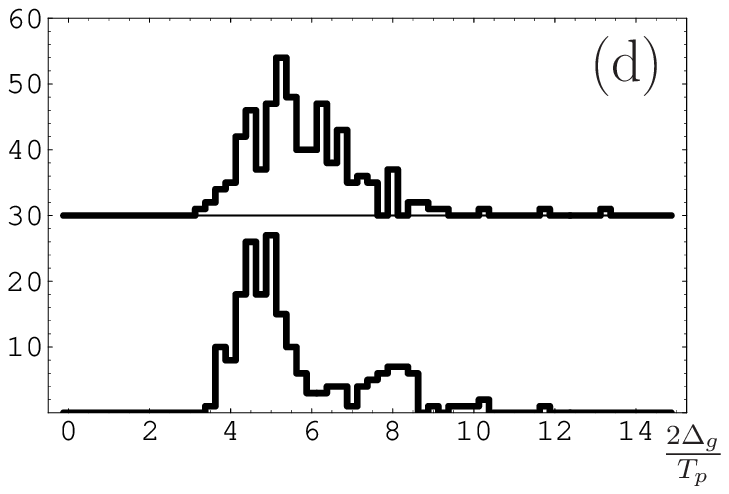}}
  \end{minipage}
    \caption{(a) The
correlation between the OP spatial distributions in models (B) and
(B'). Each point with the coordinates $(\Delta^{(B)},
\Delta^{(B')})$ represents the values of the superconducting OP
for a given site $i$ in models (B) and (B'). The line is the
linear fit $\Delta^{(B')}=1.01 \Delta^{(B)}$ to these points. (b)
Analogous correlation between the low-temperature ($T=0.03$) gaps
in the considered models. The points are fitted by the liner
dependence $\Delta_g^{(B')}=0.69 \Delta_g^{(B)} + 0.18$. (c) The
correlation between $T_p$. The linear fit is $T_p^{(B')}=1.42
T_p^{(B)} - 0.06$. (d) The histograms for $2\Delta_g/T_p$
distributions. The upper histogram corresponds to model (B'),
while the bottom is related to model (B). The offset is for
clarity.} \label{pot_dis}
\end{figure}

To investigate the influence of the discussed additional disorder
on the properties of interest here we compare the OP, gap and
$T_p$ spatial distributions for models (B) and (B'). The
correlation between the OP spatial distributions in models (B) and
(B') is demonstrated in Fig.~\ref{pot_dis}(a). The horizontal axis
represents the value of the superconducting OP in model (B), while
the vertical axis corresponds to this value for model (B'). Each
point with the coordinates $(\Delta^{(B)}, \Delta^{(B')})$
represents the values of the superconducting OP for a given site
$i$ in models (B) and (B'). The line is the linear fit
$\Delta^{(B')}=1.01 \Delta^{(B)}$ to these points. It is seen that
the points are described by this fit very well, so the additional
disorder practically does not affect the superconducting OP.

At the same time the gap and $T_p$ distributions are influenced by
the potential and hopping scatterers, as it is seen in
Figs.~\ref{pot_dis}(b) and (c), respectively. The reason is that
the gap is (at least partially) determined by the LDOS peak
originated from the spectral weight transfer from the van Hove
singularity by Andreev scattering processes, while, as it was
discussed above, the van Hove singularity is affected quite
strongly by the potential and hopping disorder. Nevertheless, the
correlations between the low-temperature gaps and $T_p$ in models
(B) and (B') can be still fitted by a linear dependence, what
indicates the fact that the corresponding gap maps do not change
qualitatively under the influence of such a disorder, but only
distorted to a certain extent. The average values of the
low-temperature gap and $T_p$ also differ very slightly for
models (B) and (B'): $\langle \Delta_g^{(B')} \rangle=0.50$ and
$\langle T_p^{(B')} \rangle=0.17$, while $\langle \Delta_g^{(B)}
\rangle=0.47$ and $\langle T_p^{(B)} \rangle=0.16$.

The distributions of the ratio $2\Delta_g/T_p$ for models (B) and
(B') are compared in Fig.~\ref{pot_dis}(d). The average value of
the ratio is again affected by the potential and hopping disorder
only slightly: $\langle 2\Delta_g/T_p \rangle=5.8$ for model (B')
and $\langle 2\Delta_g/T_p \rangle=5.6$ for model (B), although
the disorder results in widening of the distribution.

At the end of the discussion of the potential and hopping disorder
effect we would like to note that the considered here set of
parameters is just a representative example. We have studied a
number of other weak potential and hopping disorder configurations
and found that their effect qualitatively the same. Although the
particular values for the potential and hopping disorder
parameters and their relationship to the pair disorder strength
can be only established in the framework of certain microscopic
models, our analysis indicates that they do not qualitatively
influence the physics discussed in the present paper.

\section{Thermal phase fluctuations: a toy model}
\label{phase}

The considered above mean-field approximation seems to be
physically not quite appropriate for studying the state, which is
inhomogeneous on the atomic scale, especially in view of short
coherence length in cuprate materials. It is reasonable to assume
that the phase of the superconducting order parameter should
fluctuate from one region of enhanced pairing amplitude to another
in such an inhomogeneous situation, what can significantly
suppress the temperature $T_p$ and, consequently, increase the
ratio $2\Delta/T_p$. The regular consideration of the state with
inhomogeneous pairing interaction beyond the framework of
mean-field approximation, which takes into account thermal phase
fluctuations, is a separate problem. So we postpone it for a
future publication. To show that the thermal phase fluctuations
are indeed systematically suppress $T_p$, in the present paper we
study their effect in the framework of a toy model.

While in the mean-field approximation the phase of the
superconducting OP is the same over the entire sample, we assume
that it only remains constant over the region $\sim l^2$ around an
individual perturbation, where $l$ is an average distance between
the perturbations. The value of the phase in the vicinity of a
perturbation centered at a site $i$ for a given temperature $T$ is
set by hand according to the formula $\gamma T \{\Theta_i\}$,
where $\{\Theta_i\}$ is a random number belonging to the interval
$[-\pi/2,\pi/2]$ and $\gamma$ is a coefficient accounting for the
strength of the fluctuations. As it is seen from the above
expression the phases at different perturbation regions are
partially correlated for low enough temperatures and the variation
rises with temperature modelling the effect of the thermal
fluctuations. The particular law of increasing the fluctuations
with rising temperature (linear in the considered model) is not
very important. Anyway, the main effect of the thermal phase
fluctuations is to suppress $T_p$ considerably. The reason for
this is that they partially destroy the proximity effect between
different perturbation regions thus shifting the physical
properties in the vicinity of a given pair scatterer to the limit
of an independent single perturbation.

\begin{figure}[!tbh]
\begin{minipage}[b]{\linewidth}
   \centerline{\includegraphics[clip=true,width=1.7in]{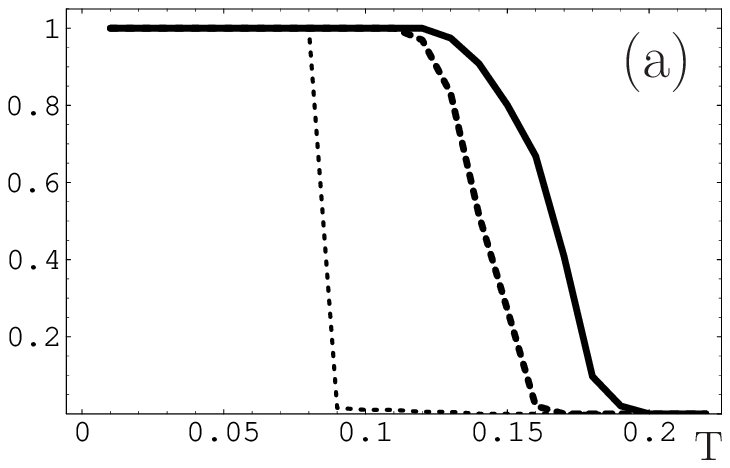}}
  \end{minipage}\hfill
 \begin{minipage}[b]{\linewidth}
   \centerline{\includegraphics[clip=true,width=1.7in]{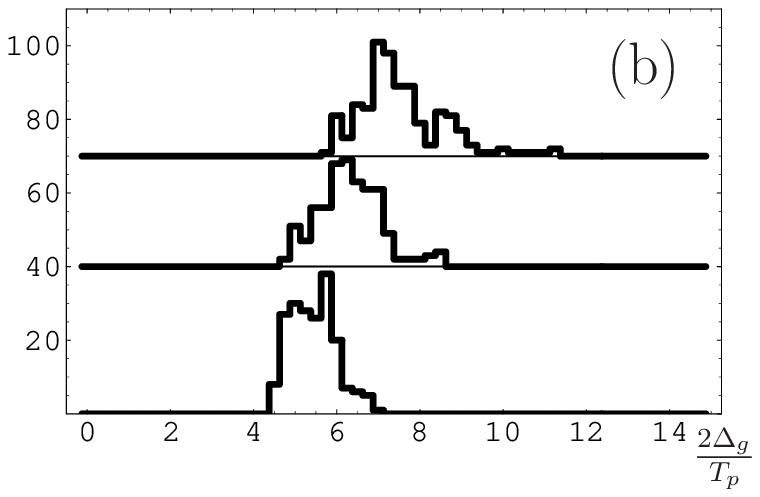}}
  \end{minipage}
    \caption{(a) The part of the sample gapped at a given temperature as a
function of temperature. The black solid line corresponds to the
mean-field results, the dashed line demonstrates the effect of
weak enough fluctuations described by $\gamma=5$ and the dotted
line is related to more stronger fluctuations with $\gamma=10$.
(b) The influence of the fluctuations on the $2\Delta_g(i)/T_p(i)$
distribution. The bottom histogram is calculated for the
mean-field sample, while the middle one is for the weaker
fluctuations with $\gamma=5$ and the upper distribution
corresponds to $\gamma=10$. All the results are related to model
(A).} \label{fluct_1}
\end{figure}

\begin{figure}[!tbh]
\begin{minipage}[b]{\linewidth}
   \centerline{\includegraphics[clip=true,width=1.7in]{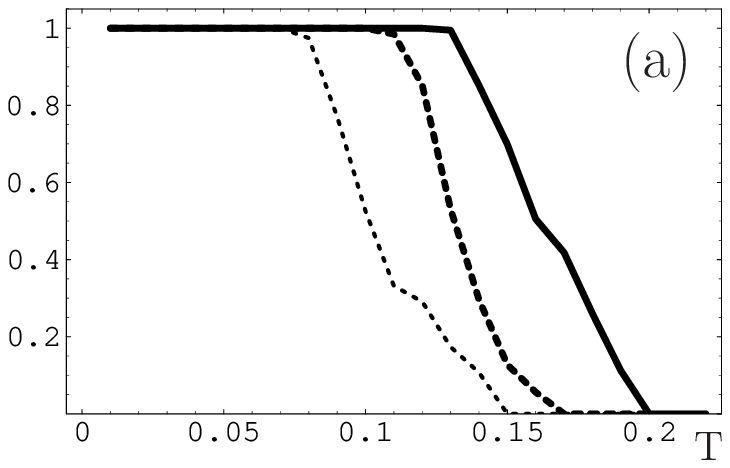}}
  \end{minipage}\hfill
 \begin{minipage}[b]{\linewidth}
   \centerline{\includegraphics[clip=true,width=1.7in]{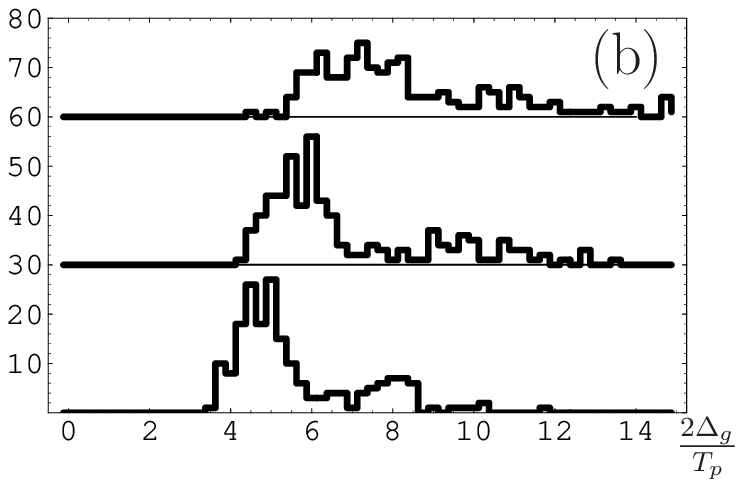}}
  \end{minipage}
    \caption{The same results as in Fig.~\ref{fluct_1},
    but calculated for model (B).} \label{fluct_2}
\end{figure}

To demonstrate the discussed above effect in the framework of our
toy model, in Figs.~\ref{fluct_1}(a) and \ref{fluct_2}(a) we
plotted the part of the sample gapped at a given temperature as a
function of temperature for models (A) and (B), respectively. The
black solid line in Fig.~\ref{fluct_1}(a) [\ref{fluct_2}(a)]
represents the results for the pure mean-field model (A) [(B)]
without phase fluctuations, the dashed lines demonstrate the
effect of weak enough fluctuations corresponding to $\gamma=5$ and
the dotted lines are related to more stronger fluctuations with
$\gamma=10$. It is seen that the increase in the fluctuation
strength $\gamma$ monotonously suppresses $T_p$.

Figs.~\ref{fluct_1}(b) and \ref{fluct_2}(b) represent the
influence of the fluctuations on the $2\Delta_g/T_p$ distributions
for models (A) and (B), respectively. The bottom histograms
correspond to the mean-field samples, while the middle ones are
related to the weaker fluctuations with $\gamma=5$ and the upper
distributions are calculated for the more stronger fluctuations
modelling by $\gamma=10$. It is clearly seen for the both models
that the increase of the phase variation monotonously shifts the
distribution to more higher values without considerable changing
of its shape. The average values of the ratio $2\Delta_g/T_p$
corresponding to all the histograms are represented in the
Table~\ref{tab:average_ratios}. It is worth to note that the
average ratios corresponding to the stronger fluctuations with
$\gamma=10$ (last column in Tab.~\ref{tab:average_ratios}) are in
excellent agreement with the appropriate ratios for the single
perturbations. This points to the fact that these fluctuations
suppress the proximity effect between neighbor enhanced pairing
regions quite efficiently, so that the limit of single
perturbation is practically reached for an individual perturbation
region.
\begin{table}
\begin{ruledtabular}
\begin{tabular}{|c|c|c|c|}
\hline $\gamma$ & 0 & 5 & 10 \\\hline (A) & 5.4 & 6.5 & 7.6
\\\hline (B) & 5.6 & 6.9 & 8.6 \\\hline
\end{tabular}
\end{ruledtabular}
\caption{The values of the ratio $\langle 2\Delta_g/T_p \rangle$
for all the histograms shown in Figs.~\ref{fluct_1}(b) and
\ref{fluct_2}(b). The first column corresponds to the mean-field
results, while the second and third ones represent their shift
under the influence of weak and strong fluctuations, respectively.
The middle row is related to model (A), and correspondingly to
Fig.~\ref{fluct_1}(b). The bottom one contains the average ratios
for model (B) and, consequently, for the histograms in
Fig.~\ref{fluct_2}(b).} \label{tab:average_ratios}
\end{table}

The results demonstrated in Figs.~\ref{fluct_1}, \ref{fluct_2} and
Tab.~\ref{tab:average_ratios} are calculated for a given random
realization of the parameters $\{ \Theta_i \}$. Having studied a
number of random realizations of these parameters we found that
the absolute error in determining the average ratio $\langle
2\Delta_g/T_p \rangle$ is $\sim 0.2$.

\section{Anticorrelation between low-temperature gap and high-temperature zero-bias conductance}
\label{anticorr_phys}

As it was shown above, one of the characteristic features of the
model with the atomic-scale inhomogeneity of the superconducting
OP is the considerable enhancement of the ratio $2\Delta_g/T_p$ as
compared to the homogeneous case. In this section we would like to
discuss another characteristic manifestation of the OP
inhomogeneity in the spectra measured by STM. It was recently
reported\cite{yazdani08} that the value of the low-temperature
position-dependent gap obtained from the STM spectra strongly
anticorrelated with the value of zero-bias differential
conductance at a temperature, where practically the entire sample
is ungapped. One of possible explanations for this experimental
observation can be naturally given in the framework of the
inhomogeneous OP model.

\begin{figure}[!tbh]
\begin{minipage}[b]{\linewidth}
   \centerline{\includegraphics[clip=true,width=1.7in]{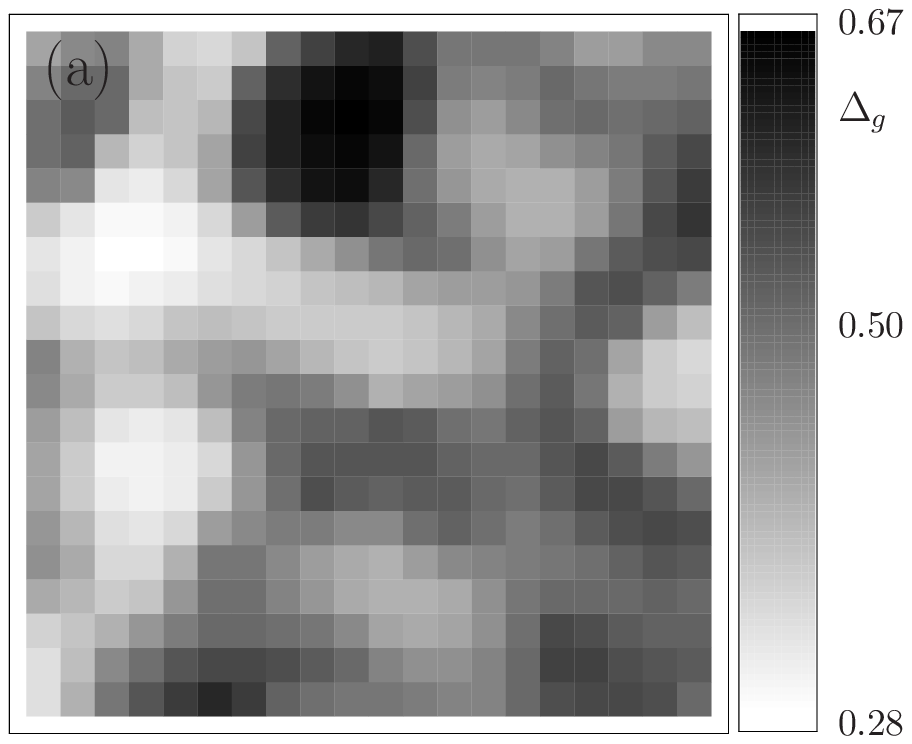}}
  \end{minipage}\hfill
 \begin{minipage}[b]{\linewidth}
   \centerline{\includegraphics[clip=true,width=1.7in]{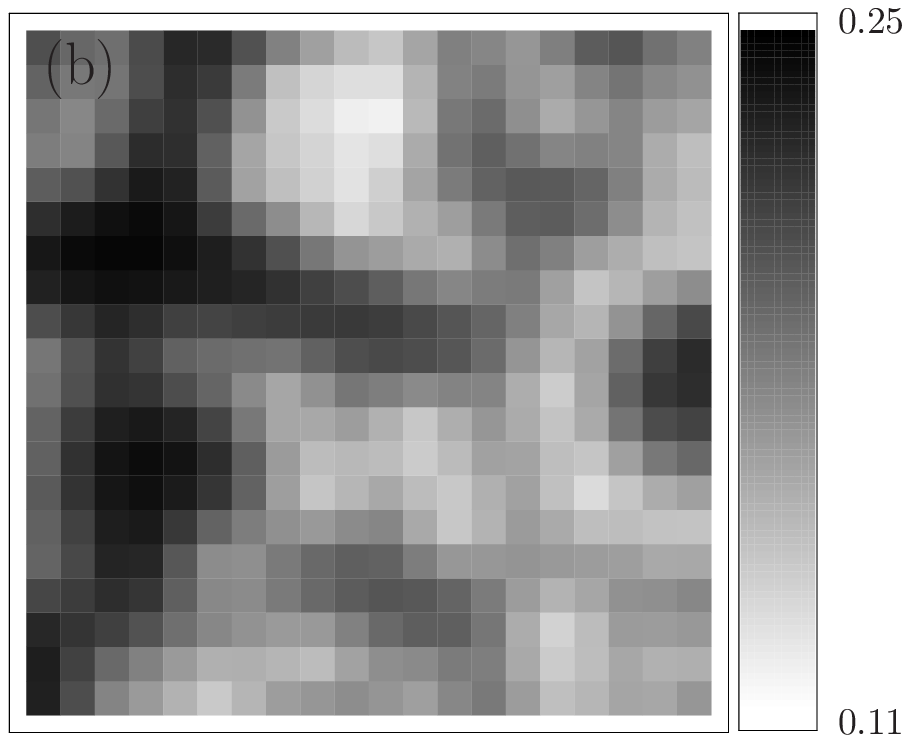}}
  \end{minipage}\hfill
 \begin{minipage}[b]{\linewidth}
   \centerline{\includegraphics[clip=true,width=1.7in]{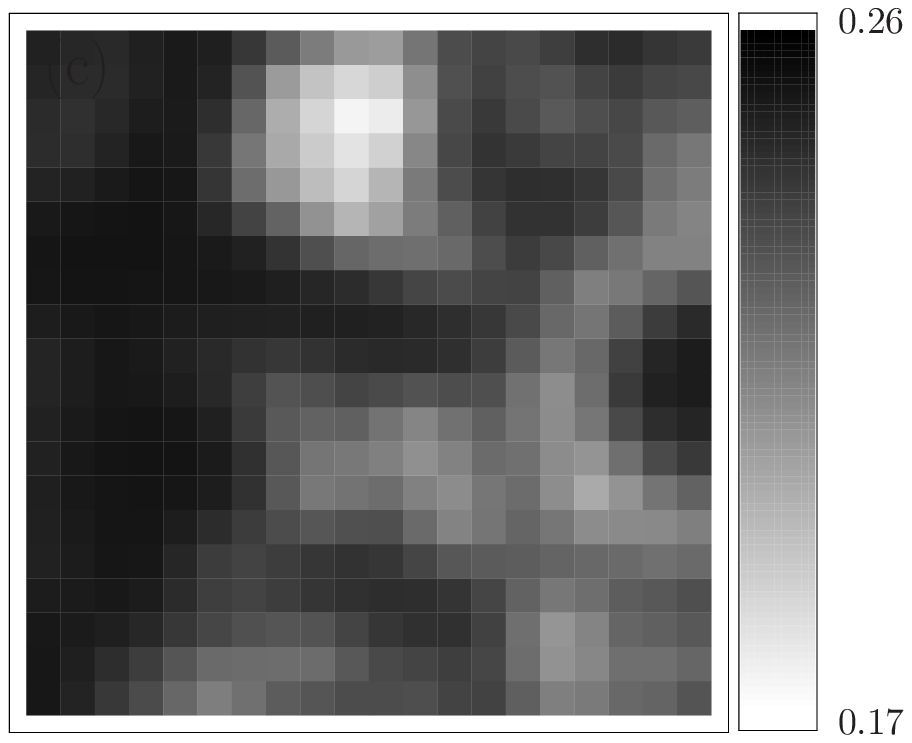}}
  \end{minipage}\hfill
 \begin{minipage}[b]{\linewidth}
   \centerline{\includegraphics[clip=true,width=1.7in]{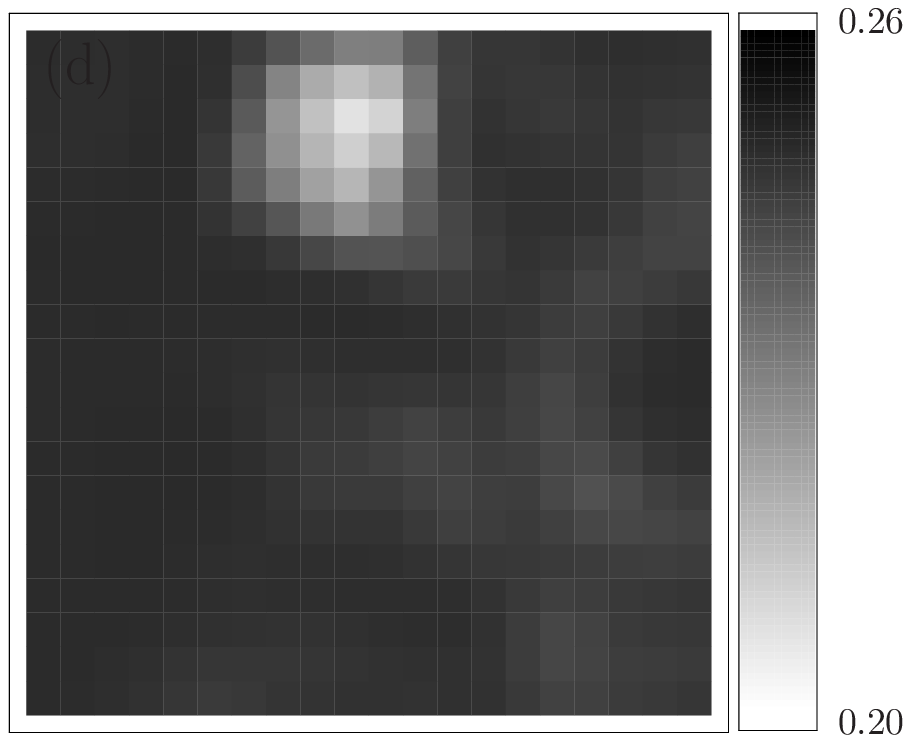}}
  \end{minipage}
    \caption{(a) The
low-temperature ($T=0.05$) gap map for model (A). (b)-(d) The
zero-bias space distributions of the thermally-smeared LDOS for
the same model: (b) $T=0.15$, (c) $T=0.19$ and (d) $T=0.21$.}
\label{correlation}
\end{figure}

\begin{figure}[!tbh]
\begin{minipage}[b]{\linewidth}
   \centerline{\includegraphics[clip=true,width=1.7in]{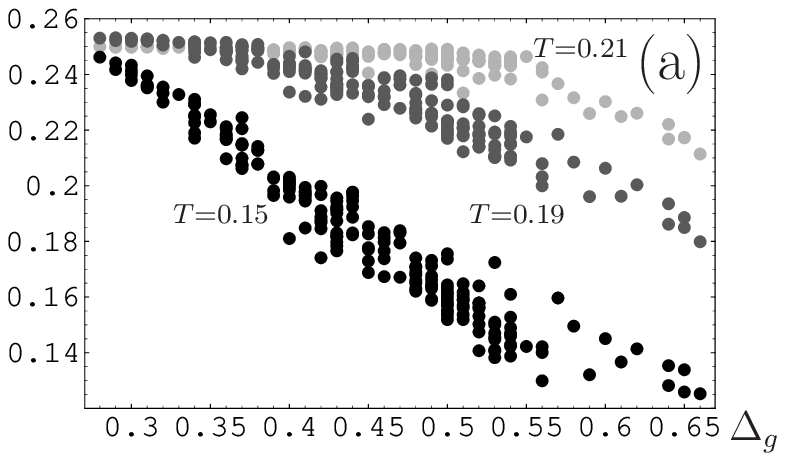}}
  \end{minipage}\hfill
 \begin{minipage}[b]{\linewidth}
   \centerline{\includegraphics[clip=true,width=1.7in]{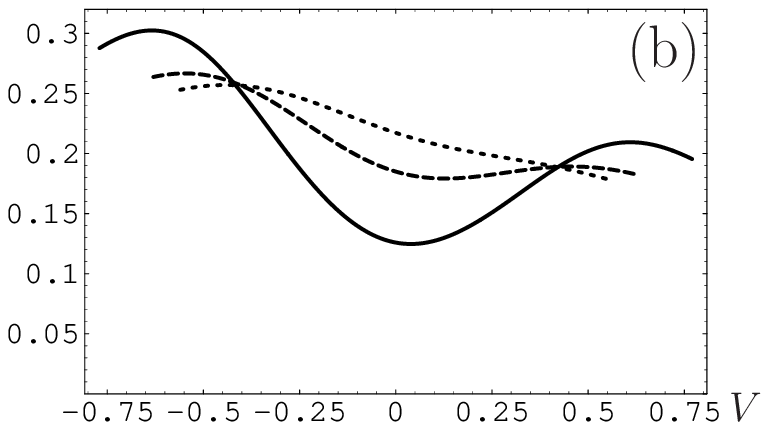}}
  \end{minipage}\hfill
 \begin{minipage}[b]{\linewidth}
   \centerline{\includegraphics[clip=true,width=1.7in]{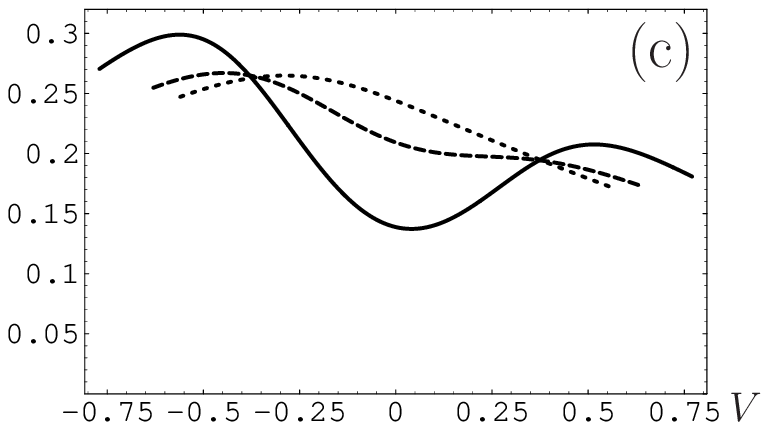}}
  \end{minipage}\hfill
 \begin{minipage}[b]{\linewidth}
   \centerline{\includegraphics[clip=true,width=1.7in]{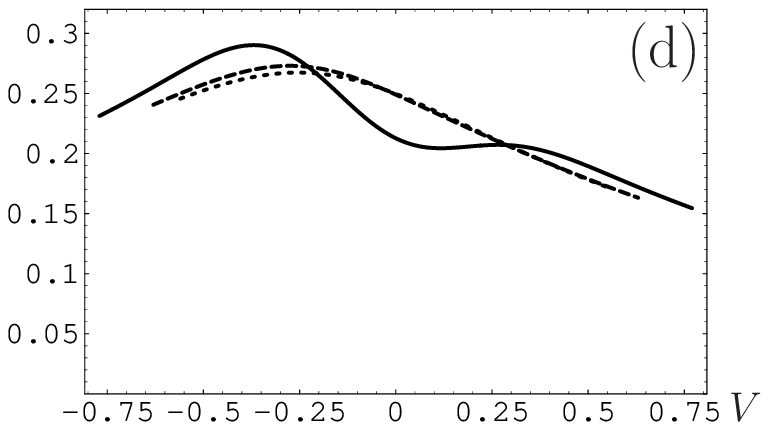}}
  \end{minipage}
    \caption{(a) High-temperature zero-bias thermally-smeared LDOS, measured
along the vertical axis, versus low-temperature gap (horizontal
axis) for all the sites of the sample at three different
temperatures: $T=0.15$, $T=0.19$ and $T=0.21$. (b)-(d) The
temperature evolution of thermally smeared LDOS for three typical
spatial locations. $T=0.15$ (solid black lines), $T=0.19$ (dashed
lines) and $T=0.21$ (dotted lines). All the panels are related to
model (A).} \label{high_T_spectra}
\end{figure}

Fig.~\ref{correlation} demonstrates the low-temperature gap-map
(panel (a)) in comparison with the maps of high-temperature
zero-bias thermally smeared LDOS (panels (b)-(d)). All the maps
correspond to the considered above model (A) in the mean-field
approximation. High-temperature zero-bias thermally smeared LDOS
is calculated according to the formula (\ref{smeared_DOS}) at V=0
for three different temperatures: the results for $T=0.15$, where
$80 \%$ of the sample are still gapped, are shown in panel (b),
while panel (c) is related to $T=0.19$, where only $2\%$ of the
sample are gapped and panel (d) corresponds to $T=0.21$, where the
entire sample is already ungapped. The anticorrelation between the
maps in panel (a) from one hand and the maps in panels (b)-(d) is
clearly seen. Fig.~\ref{high_T_spectra}(a) further quantifies this
anticorrelation. High-temperature zero-bias thermally smeared
LDOS, measured along the vertical axis, versus low-temperature gap
(horizontal axis) is plotted in this figure for all the sites of
our sample.

As it was described above in detail, the atomic-scale OP
inhomogeneity leads to the fact, that the difference between the
temperature of vanishing the gap $T_p(i)$ and the temperature of
disappearing the pair correlations, which corresponds to the
critical temperature $T_x$ in the mean-field treatment, is
considerably higher for the inhomogeneous situation as compared to
the homogeneous case. The temperature evolution of thermally
smeared LDOS from $T=0.15$ to $T=0.21$ is demonstrated in
Figs.~\ref{high_T_spectra}(b)-(d) for three typical space
locations. Only very small part of the sample exhibits the
behavior shown in panel (b), while approximately $2/3$ of the
sample can be described by panel (c) and the behavior of the
conductance for the last $1/3$ of the sites corresponds to panel
(d).

That is, although practically the entire sample is ungapped at
$T=0.19$ for our model (A) and the conductance curves change only
slightly or practically do not change under further increasing in
the temperature, the value of the superconducting OP is not small
for the most part of the sample at this temperature (see
Fig.~\ref{Delta_T_many}(a)). Even the complete vanishing of the
gap at $T=0.21$ does not mean that there are no pair correlations
in the system. The presence and the strength of the pair
correlations is reflected in the suppression of the low-energy
LDOS. It is clearly seen in Figs.~\ref{correlation}(a)-(d) and
\ref{high_T_spectra}(a) that the anticorrelation weakens with
temperature. It is rather weak at $T=0.21$, where the pair
correlations are small, and should disappear in the framework of
our model when the pair correlations entirely vanish. We believe
that this scenario should also work if one takes into account the
thermal phase fluctuations because the same physical picture is
valid for the single perturbation as well. However, a regular
theory is needed for a detailed consideration of this problem.

\section{Conclusions}
\label{conclusions}

We have studied the influence of the atomic-scale inhomogeneities
of the superconducting OP on the conductance spectra measured by
STM. First of all, it is found that the ratio of the local
low-temperature gap in differential conductance spectra to the
local temperature of vanishing the gap $2\Delta_g/T_p$ can take
large enough values as compared to the homogeneous OP model. While
in the framework of the mean-field approximation the ratio does
not strongly differ from the homogeneous one, the thermal phase
fluctuations considerably enhance it. At least in the framework of
a very simplified model we obtained that the ratio $2\Delta_g/T_p$
can reach the values $\sim 7-8$, which are comparable to the
experimental ones. It is also demonstrated that the additional
weak potential scatterers and hopping matrix element disorder do
not influence qualitatively the above results. Second, the
atomic-scale OP inhomogeneity results in the anticorrelation
between the low-temperature gap and the high-temperature zero-bias
conductance, bearing a resemblance to the recent results obtained
by STM.

\section{Acknowledgments}

The support by the Russian Science Support Foundation (A.M.B.), RF
Presidential Grant No.MK-4605.2007.2 (I.V.B.) and the programs of
Physical Science Division of RAS are acknowledged.

\end{document}